\documentclass[twocolumn]{revtex4}
\usepackage{graphicx}

\newcommand{\be}{\begin{equation}}
\newcommand{\ee}{\end{equation}}
\newcommand{\ra}{\rangle}
\newcommand{\la}{\langle}

\newcommand{\bt}{\beta}

\newcommand{\al}{\alpha}

\begin{document}

\title{The Piecewise Moments Method: A Generalized Lanczos Technique for Nuclear Response Surfaces}

\author{Wick C. Haxton}
\affiliation{Institute for Nuclear Theory and Dept. of Physics, University of Washington,
  Seattle, WA 98195}
\author{Kenneth M. Nollett}
\affiliation{Physics Division, Argonne National Laboratory, Argonne, IL  60439}
\author{Kathryn M. Zurek}
\affiliation{Institute for Nuclear Theory and Dept. of Physics, University of Washington,
  Seattle, WA 98195}

\begin{abstract}
For some years Lanczos moments methods  have been combined
with large-scale shell-model calculations in evaluations of the
spectral distributions of certain operators.   This technique is of great value because
the alternative, a state-by-state summation over final states,
is generally not feasible.  The most celebrated application is to the
Gamow-Teller operator, which governs $\beta$ decay and neutrino
reactions in the allowed limit.  The Lanczos procedure determines
the nuclear response along a line $q$=0 in the $(\omega,q)$
plane, where $\omega$ and $q$ are the energy and three-momentum
transfered to the nucleus.  However, generalizing such treatments
from the allowed limit to general electroweak
response functions at arbitrary momentum transfers
seems considerably more difficult: the response function must be
determined over the entire $(\omega,q)$ plane for an operator
$O(q)$ that is not fixed, but depends explicitly
on $q$.   Such operators arise in any semileptonic process where 
the momentum transfer is comparable to (or larger than) the inverse nuclear size.
Here we show, for Slater determinants built on harmonic oscillator basis functions,
that the nuclear response for any multipole operator $O(q)$ can be determined 
efficiently over the full response plane by a generalization of the
standard Lanczos moments method.
We describe the Piecewise Moments Method and thoroughly explore its
convergence properties for the test case of electromagnetic responses in a
full $sd$-shell calculation of $^{28}$Si.  We discuss possible extensions to a 
variety of electroweak processes, including charged- and
neutral-current neutrino scattering.
\end{abstract}

\date{\today}
\maketitle

\section{Introduction and Motivation}

The shell model \cite{caur} is an important tool for modeling the ground
states and low-lying excitations of nuclei, as well as their
electromagnetic and weak-interaction properties.  Calculations
are performed by diagonalizing  an effective interaction $H$
within a low-momentum model space typically consisting of
all Slater determinants that can be constructed within one or perhaps a
few principal shells.  $H$ corrects for the absence of the
remaining shells and of the high-momentum correlations that
reside primarily in those excluded shells.  Typically $H$
is determined empirically \cite{brown}, though recently considerable work
has been invested in various effective-theory approaches to
derive $H$ directly from the underlying bare interaction \cite{song}.

With increasing CPU speeds and the advent of parallel
computing, shell model practitioners have been able to tackle
problems in very large bases -- ranging in some cases to Hamiltonians
of dimension $\sim 10^8 - 10^9$ \cite{caur}.   Because direct diagonalizations in
such spaces are impossible, iterative methods are used to
determine extremum eigenvalues and their eigenfunctions.
The Lanczos algorithm \cite{lanczos} is the most commonly used method.
It is based on a recursive mapping of the full Hamiltonian
into a much smaller, tridiagonal matrix that preserves certain
exact information on the lowest moments of
the full Hamiltonian.   As described in the next section,
extremum eigenvalues of the Lanczos matrix converge to those of the full 
Hamiltonian because of the algorithm's moments properties,
thus making the algorithm useful as a diagonalization
tool.  However, its real power is connected with another
property, a stable method for solving the classical moments
problem -- determining the simplest discrete distribution
characterized by those same lowest moments. 

One can argue that the Lanczos
algorithm adresses both of the most common challenges encountered
in nuclear structure physics -- detailed information
about specific low-lying states and global information on
spectral moments important to inclusive properties, such as
response functions, polarizabilities, and Green's functions.

The moments method, described in
more detail in the next section, can be applied in a straight-forward
way to operators like the Gamow-Teller (GT) operator,
\[ \sum_{i=1}^A \sigma(i) \tau_{\pm}(i). \]
This operator is independent of  the
three-momentum transfer $q$.  GT strength distributions
(and their neutral-current analogs) govern low-energy neutrino
reactions important to solar and supernova physics.  The combination 
of increasingly sophisticated shell-model ground-state wave functions
and Lanczos methods for strength distributions has had great impact.
Results include much better agreement
between theory and experiment (e.g., $(p,n)$ mappings of GT
distributions) and new iron-group weak interaction rates \cite{langanke00}
that have influenced the collapse and explosion physics
of Type II supernovae
\cite{heger01,hix03}.  

However, the restriction to allowed operators is very limiting.
For example, in supernova physics, approximately half of the strength for heavy-flavor
neutral current scattering is carried by momentum-dependent operators \cite{haxton88}.  
Because no efficient moments method is available for these more complex operators,
and because state-by-state summations would be prohibitively difficult, theorists
have not been able to make use of state-of-the-art shell model wave functions
in evaluating inclusive forbidden response functions.  Instead less sophisticated methods --
simplified shell model treatments, QRPA, or even schematic approaches
like the Goldhaber-Teller model -- have been substituted, as the resulting
smaller Hilbert spaces do allow state-by-state summations.
The difficulty in designing a moments method 
for a multipole operator $O(q)$ that varies with $q$ is clear: the standard moments method
would require one to repeat the calculation many times over a grid in 
$q$, as $O(q)$ is a fixed operator only along $q$=constant trajectories.
Furthermore, as will become clearer below, a rather dense grid in $q$ would be needed
to interpret an experiment that maps out some nontrivial trajectory in the
$(\omega,q)$ response plane, as the discreteness of the moments distributions
at each $q$ makes it complicated to extrapolate in $q$. 

Interest in general electroweak response functions
is not limited to supernova physics, clearly.  An electron
scattering experiment, for example, generally maps out some area within
the space-like half of the response plane, depending on the range of
spectrometer angles and electron energy loss explored. 

In this paper we show that there is an efficient moments technique 
for constructing the entire shell-model inclusive response surface -- the response 
$S(\omega,q)$ as a continuous function of both $q$ and $\omega$, as well as of the
oscillator parameter $b$ --
provided the underlying single-particle basis wave functions are taken to
be harmonic oscillators.  The method involves a small number of Lanczos
calculations  (at most three in the $^{28}$Si test case we use here).  That is, the full
response surface can be reconstructed with little more effort than required in the
familiar GT case, where results are obtained only for a line $q$=0
in the response plane.  The information that must be extracted from the many-body
calculations to perform this reconstruction are the elements of the tridiagonal
Lanczos matrices and certain dot products between Lanczos vectors.
The method can be viewed as a nearly perfect ``numerical effective theory," as it extracts from potentially quite complicated
microscopic calculations precisely that information necessary to 
reconstruct the response surface up to a specified resolution in energy.
This has implications for problems like modeling supernova explosions,
where one issue in the use of sophisticated nuclear physics results is
defining a practical scheme for making the information available ``on line"
within the explosion codes.
  
The outline of this paper is as follows: In the next section we review
in more detail the Lanczos method, including its conventional uses in
constructing response functions and Green's functions.  In the third
section we discuss the problem at hand, the construction of response
functions for electroweak nuclear operators at arbitrary $(\omega,q)$.
We discuss properties of harmonic oscillator matrix elements of these
operators which could lead to efficient algorithms for constructing
the full response surface.  We describe numerical limitations to some
possible approaches.  In the fourth section we describe the Piecewise
Moments Method (PMM), which we formulate first for a Lorentzian
resolution function in energy, but then generalize for any choice.  In
the fifth section we explore convergence properties of the PMM, using
as a test case the electromagnetic response functions for a full
$sd$-shell calculation of $^{28}$Si.  Convergence properties of the
method are explored and various numerical comparisons are made
with ``exact" results.  A series of results for response surfaces
are presented.  In the concluding section we discuss opportunities for
applying the method, including supernova physics and the neutrino
reactions at energies below 1 GeV (e.g., LSND or atmospheric
neutrinos).  We suggest extensions of the work -- to check unitarity
and to project spurious states -- that might be important in future
applications of the PMM.

\section{Lanczos Algorithm Preliminaries}

The Lanczos method is based on a mapping of a
Hamiltonian $H$ of dimension $N$ to tridiagonal form by a recursive definition
of a new orthonormal basis $|v_i \ra$, called the Lanczos vectors.
One begins with an arbitrary unit vector $ |v_1\ra$
and performs the successive operations to define the $|v_i \ra$,
\begin{eqnarray}
H|v_1\ra=&\alpha_1|v_1\ra+\beta_1|v_2\ra \nonumber \\
H|v_2\ra=&\bt_1|v_1\ra+\al_2|v_2\ra+\bt_2|v_3\ra \nonumber \\
H|v_3\ra=&\bt_2|v_2\ra+\al_3|v_3\ra+\bt_3|v_4\ra \nonumber \\
H|v_i\ra=&\bt_{i-1}|v_{i-1}\ra+\al_i|v_i\ra+\bt_i|v_{i+1}\ra. \label{eqn:lanczos}
\end{eqnarray} 
where in the first step $\alpha_1|v_1\ra$ is the projection of
$H|v_1\ra$ onto $|v_1\ra$, while the remaining orthogonal portion
defines a new unit vector and amplitude $\beta_1|v_2\ra$.  In the
third step we see the tridiagonal form emerge, as the construction
demands $\la v_3| H |v_1\ra$ = 0.  The hermiticity of $H$ has been
used above.  In practice this algorithm, when used to determine
extremum eigenvalues and eigenvectors, is usually executed with a
reorthogonalization step in each iteration, to guarantee that the new
vector $|v_i\ra$ is orthogonal to all previous vectors.  While this
step is not required mathematically, it is nevertheless important
numerically, as roundoff error can grow with successive iterations
\cite{cullum85}, eventually leading to a Lanczos matrix that contains
multiple copies of subspaces of $H$.

Clearly, if the procedure were continued $N$ steps, 
one would find $\bt_N$ =0, as exhaustion of the Hilbert
space terminates the construction.  The effect of the
Lanczos procedure would be a unitary transformation to a new basis
in which the Hamiltonian is tridiagonal, but still of dimension $N$.  However,
in useful applications $N$ is very large, and the Lanczos construction
is truncated by choice after $n$ iterations by setting $\bt_n = 0$, $n \ll N$.
This yields the truncated tridiagonal Lanczos matrix $L(n,|v_1\ra)$
\begin{equation}
\label{eqn:lanczosmatrix} L(n,|v_1\ra)=\left(\begin{array}{cccccc}
\alpha_1&\beta_1& 0 & 0 & &\\ \beta_1&\alpha_2&\beta_2& 0 &&\cdots\\ 0
&\beta_2&\alpha_3&\beta_3&&\\ 0 & 0 &\beta_3&\alpha_4&&\\
&&\vdots&&\ddots&\\ &&&&\bt_{n-1}&\al_n\\
\end{array} \right)
\end{equation}
The notation emphasizes that $L$ is entirely determined by the
number of iterations $n$ and by the
choice of the starting vector $|v_1\ra$.

The powerful property of this mapping of $H$ onto a much smaller subspace
is that it extracts specific exact information from the full Hamiltonian
\begin{eqnarray}
&H(N) \rightarrow L(n,|v_1\ra)&  \nonumber \\
&\{\la v_1 | H^\lambda | v_1\ra, \lambda=1,\ldots,2n-1\} \leftrightarrow& \nonumber \\
&\{\al_1,\ldots,\al_n; \bt_1, ...,\bt_{n-1}\}&.
\end{eqnarray}
That is, the first $2n-1$ moments of H for the starting vector $|v_1\ra$ determine
$L(n,|v_1\ra)$, and conversely.  In a very real sense, the Lanczos algorithm
can be considered a numerical effective theory.  It systematically extracts from the
large matrix the long-wavelength information describing the distribution of $|v_1\ra$
over the eigenspectrum of $H$, while leaving behind the high
frequency information important to the detailed structure of this distribution, but not
to any of its broad features.  More precisely, if we denote by $\{(\psi_{E_i},E_i),i=1,\ldots,N\}$
the exact eigenenergies/functions of $H$ and by $\{(\hat{\psi}_{\hat{E}_i},\hat{E}_i),i=1,\ldots,n\}$ the
eigenenergies/functions of $L(n,|v_1\ra)$, then
\begin{eqnarray}
\label{eqn:strength}
\la v_1 | H^\lambda | v_1 \ra &=& \sum_{i=1}^N |\la v_1 | \psi_{E_i} \ra|^2 E_i^\lambda \nonumber \\
&= &\sum_{i=1}^n |\la v_1 | \hat{\psi}_{\hat{E}_i} \ra|^2 \hat{E}_i^\lambda \nonumber \\
& =&\sum_{i=1}^n |\hat{\psi}_{\hat{E}_i}(1)|^2  \hat{E}_i^\lambda, \lambda=1,\ldots,2n-1 
\end{eqnarray}
That is, the eigenvalues and eigenvectors of the Lanczos matrix $L(n,|v_1\ra)$ determines a set of
$n$ points $\hat{E}_i$ and weights that solves the classical moments problem --
a discrete distribution whose moments reproduce those of the full matrix $H$.  The
weights are simply the squares of the first components of the respective Lanczos 
eigenvectors.  As Whitehead has emphased \cite{whit1,whit2}, the speed and numerical
stability of this classical moments solution is a very special property of the Lanczos
algorithm.

The usual application of the Lanczos algorithm is in determining extremum
eigenvalues, especially the ground state and first few excited states.  After each
iteration $L(n,|v_1\ra)$ can be diagonalized by standard techniques, determining
eigenvalues and eigenfunctions.  Because extremum eigenvalues are heavily
weighted in $H^n$ when $n$ is large and often separated by gaps from the bulk
of spectrum, these eigenvalues of $L$ must quickly converge to the true eigenvalues
of $H$.  This convergence can be monitored numerically as the algorithm is
executed.  In typical shell model applications the ground state often converges in
about 50 iterations, while the lowest 10 or so eigenvalues and eigenfunctions
may be resolved in 200 iterations \cite{whit2}.

But perhaps the most elegant applications of the Lanczos algorithm are in 
distribution functions or Green's functions, inclusive quantities that depend 
most directly on spectral moments,
the long-wavelength information extracted from $H$.  Consider the response 
function $S(\omega)$
\begin{equation}
\label{eqn:responsefunction}
S(\omega) = \sum_{i=1}^N |\la \psi_{E_i} | O |g.s. \ra|^2 \delta(\omega-E_i)
\end{equation}
where the sum extends over a complete set of states $i$ of the full 
Hamiltonian $H$.  Assume for the moment that $O$ is a fixed operator,
like the Gamow-Teller operator, independent of the three-momentum
transfer $q$ to the nucleus.  We define a unit vector $|v_1\ra$ by
\begin{equation}
\label{eqn:vector}
O |g.s. \ra \equiv c |v_1 \ra,
\end{equation}
where $c$ is the overall strength (norm).  Taking
$|v_1 \ra$ as the starting vector and competing $n$ Lanczos iterations,
one can form the distribution
\begin{equation}
\label{eqn:strengthdistribution}
\hat{S}_n(\omega) = |c|^2 \sum_{i=1}^n \delta(\omega-\hat{E}_i) |\hat{\psi}_{\hat{E}_i}(1)|^2.
\end{equation} 
If one weights this distribution with $\omega^\lambda$ and integrates over $\omega$, 
one sees from Eq.~\ref{eqn:strength} that $S_n(\omega)$ reproduces
the lowest $2n-1$ integrated moments of the exact spectral distribution
$S(\omega)$.

Now, as experiments are done with finite resolution, 
spectral variations occurring at energy scales below that resolution are irrelevant.  
Furthermore, the discrete spectrum of the shell model is itself an artifact of
the use of a finite Hilbert space.  It represents resonances in the continuum by 
discrete doorway states, with the density of such states increasing as
the shell-model space is expanded, to better represent the continuum.
Thus one quickly recognizes that the high-frequency information missing 
from Eq.~\ref{eqn:strengthdistribution} may be irrelevant when comparing
with experiment.  It is customary to replace the $\delta$-function in the Lanczos strength 
distribution with a resolution function, choosing a width parameter $\sigma$
appropriate to the experiment in question 
\begin{equation}
 \label{eqn:resolution}
\hat{S}_n(\omega,\sigma) = c^2 \sum_i^n R(\omega-\hat{E}_i,\sigma) |\hat{\psi}_{\hat{E}_i}(1)|^2.
\end{equation}
where $R$ is normalized to 1.  Common choices are Lorentzians and 
Gaussians, e.g.,
\begin{eqnarray}
R_L(\omega-\hat{E}_i,\sigma) &=& {\sigma \over \pi} {1 \over (\omega-\hat{E}_i)^2 + \sigma^2} \nonumber \label{eqn:res1} \\
R_G(\omega-\hat{E}_i,\sigma)&=& {1 \over \sigma \sqrt{2 \pi}} \exp{(-(\omega-\hat{E}_i)^2/2 \sigma)}
\label{eqn:res2}
\end{eqnarray}
Because such smoothing makes high-frequency information irrelevant, the Lanczos
approximation $\hat{S}_n(\omega,\sigma)$ {\it converges} to the exact smoothed distribution
$S(\omega,\sigma)$ for $n$ sufficiently large.  Qualitatively, convergence is achieved
when the  typical separation in energy of
neighboring Lanczos eigenvectors becomes substantially smaller than the chosen 
resolution $\sigma$.  For $\sigma \sim 0.25$ MeV, a value typical of (p,n) mappings 
of GT strength, for instance, this may occur at $n \sim 200$.  Thus the Lanczos 
algorithm provides an exact method for determining appropriately smoothed
strength functions in very complex shell-model spaces.

A second spectral application is to Green's functions \cite{haydock}, which arise in nuclear
calculations of polarizabilities, in virtual processes like double $\beta$ decay, and
in a variety of many-body applications, such as effective interactions and operators.
The Green's function acting on a normalized vector $|v_1\ra$
\begin{equation}
G(\omega) |v_1\ra = {1 \over \omega - H} |v_1\ra
\end{equation}
can be approximated after $n$ Lanczos iterations as
\begin{equation}
\hat{G}_n(\omega) |v_1\ra =
\hat{g}_1(\omega) |v_1\ra + \hat{g}_2(\omega) |v_2\ra + \cdots + \hat{g}_n(\omega) |v_n\ra \label{eqn:greens}
\end{equation}
where the coefficients $\hat{g}_i(\omega)$ are finite continued 
fractions formed from the entries in the tridiagonal matrix.  For example
\begin{eqnarray}
\hat{g}_1(\omega) &=& {1 \over \omega - \alpha_1 -{\beta_1^2 \over \omega - \alpha_2 -
{\beta_2^2 \over \omega - \alpha_3 - \beta_3^2}}}  \nonumber \\
&& ~~~~~~~~~~~~~~~~~~~~~~~~~~~~~~~~~~~~~\ddots
\end{eqnarray}
With each additional iteration, one additional Lanczos vector is added
to the expansion, and each continued fraction increases in rank by one
through the addition of a new $\alpha_{n+1}$ and $\beta_n$.  As most
Green's function applications involve convolutions with relatively
smooth operators, often $\hat{G}_n(\omega) |v_1\ra$ becomes
numerically equivalent to $G(\omega) |v_1 \ra$ after a few Lanczos
iterations ($\sim$ 20) \cite{engel}.

An important consequence of Eq.~\ref{eqn:greens} is that, once the
Lanczos calculation is completed, the Green's function is known as a {\it function}
of $\omega$.  This will be important in the applications we discuss later.

\section{Electroweak Response Functions at Arbitrary $q^2$}

The discussion of the previous section addressed the special case
of a fixed operator, like the GT operator, which 
governs the weak nuclear response along the $q$=0 line in
the $(\omega,q)$ response plane.  However many electroweak processes of
interest -- intermediate-energy electron or neutrino scattering, muon capture,
etc. -- involve appreciable three-momentum transfers (and 
the associated excitation of radial modes in the nucleus).  That is,
the relevant response function is
\begin{equation} 
\label{eqn:responsefunctionx} 
S(\omega,q) = \sum_{i=1}^N |\la \psi_{E_i} | O(q) |g.s. \ra|^2 \delta(\omega-E_i)
\end{equation}
where $O(q)$ is (an assumed one-body) electroweak operator that
depends explicitly on $q$.  If one naively applies the formalism
of the preceding section, a new calculation
would be needed for each desired $q$, because the operator evolves
with $q$.  This would require tediously  stepping over a 
grid of fixed $q$s, computing a Lanczos calculation for each value,
to map the full surface above the response plane.

Here we discuss procedures for evaluating $S(\omega,q)$ very 
efficiently as a function of $q$ (and $\omega$) over the entire
response plane, at the cost of only a few Lanczos calculations.
The approach depends on the assumption that the shell-model basis of
Slater determinants has been formed from harmonic
oscillator single-particle wave functions.  This choice allows one to
exploit attractive properties of the matrix elements of $O(q)$ between 
such wave functions.

While we will delay details of the test application (electromagnetic
response functions for $^{28}$Si) to the next section, here we sketch
the basic idea.  One can write $O(q) |g.s.\ra$ in second quantization
\begin{equation}
\label{eqn:sq}
\sum_{\alpha,\beta} \la \alpha | O(q) | \beta \ra a_\alpha^\dagger a_\beta |g.s.\ra .
\end{equation}
where $\alpha$ and $\beta$ represent a complete set of single-particle
quantum numbers.  For the choice of harmonic oscillators,
matrix elements 
of the standard charge, longitudinal, transverse electric,
and transverse magnetic multipoles can be evaluated in closed
form, leading to \cite{walecka,donnelly}
\begin{equation}
\la \alpha | O_J(q) | \beta \ra = y^{(J-K)/2} e^{-y} p^{\alpha \beta}(y).
\end{equation}
Here we have denoted the multipolarity of the operator by J, K=2(1) for
normal(abnormal) parity operators, and $y=(qb/2)^2$ with $b$ the
oscillator parameter.  The crucial point is that $p(y)$ is a finite polynomial
in $y$ or $q^2$.  In the $^{28}$Si test case, the most complicated operator
that arises has only three nonzero terms in $p(y)$.

We first go through a schematic argument to show how this $y$ dependence
might be exploited.  Denoting the order of the polynomial $p$ by $m$, it follows that
\begin{eqnarray}
&O(q) |g.s.\ra = & \nonumber \\
&y^{(J-K)/2} e^{-y} [c_0 |v_1^0 \ra + c_1 y |v_1^1 \ra
+ \cdots + c_m y^m |v_1^m \ra]~~~~~& \nonumber \\
&\equiv y^{(J-K)/2} e^{-y} c(y) |v_1(y) \ra&
\end{eqnarray}
using a notation analogous to Eq.~\ref{eqn:vector}, with the strength $c_j$
chosen to make $|v_1^j \ra$ a unit vector.  For parity-conserving
interactions and standard phase conventions, all quantities can be taken
as real, with the $c$s nonnegative.  The $| v_1^j \ra$, of course, are not
orthonormal.  Similarly $c(y)$ and $|v_1(y) \ra$ can be viewed as a $y$-dependent
strength and unit vector.  It follows that
\begin{equation}
\label{eqn:s}
S(\omega,q) = y^{J-K} e^{-2y} |c(y)|^2 \sum_{i=1}^N |\la \psi_{E_i} | v_1(y) \ra|^2
\delta(\omega-E_i)
\end{equation}
where
\begin{equation}
\label{eqn:spect1}
|c(y)|^2  |\la \psi_{E_i} | v_1(y) \ra|^2 = \sum_{j,k=0}^m c_k^* c_j y^{j+k}
\la v_1^k | \psi_{E_i} \ra \la \psi_{E_i} |v_1^j \ra,
\end{equation}
so that the response function has a similar polynomial form.  It also
follows that moments of $S(\omega)$ have the form
\begin{equation}
\label{eqn:spect2}
\int_0^\infty S(\omega) \omega^\lambda d\omega=
y^{J-K} e^{-2y} |c(y)|^2 \sum_{i=1}^N |\la \psi_{E_i} | v_1(y) \ra|^2 E_i^\lambda.
\end{equation}
These last two results simply state that if one had a complete set
of $N$ eigenvalues and eigenfunctions, each contributing transition
probability would have a simple, analytical behavior in $y$. 

Of course, these results are only of academic interest: as we are assuming $N$ is
prohibitively large, a complete diagonalization is impossible.
This leaves a much more interesting question: can we find an analog
of Eq.~\ref{eqn:strengthdistribution} or \ref{eqn:resolution}, an efficient
Lanczos representation of $S(\omega,q)$, that also exploits the polynomial
behavior of the response in $y$?  If so, it would appear to be a practical
way to construct the response over the entire $(\omega,q)$ plane.

We have explored several of the possibilities,
uncovering some of the numerical pitfalls.  Even the less
successful methods are interesting conceptually, so we 
describe the approaches qualitatively below, reserving details
for the Appendix.  Finally, we describe and test the Piecewise
Moments Method, demonstrating that it is effectively exact and
stable for an arbitrary number of iterations. 

\textbf {The Naive Moments Method:} The most straightforward
approach to the above problem exploits the equivalence between
the tridiagonal matrix $L(n,|v_1(y)\ra)$ and the moments
$\la v_1(y) | H^\lambda | v_1(y) \ra, \lambda=1,\dots,2n-1$.
If we had a method to determine $L(n,|v_1(y)\ra)$ as a continuous function of $y$, 
Eq.~\ref{eqn:strengthdistribution} could be then be used at any $y$.
The result would be a discrete distribution along any line $y$=constant,
that would evolve smoothly as $y$ is changed.

The moments equivalence shows this is possible as
\begin{equation}
\la v_1(y) | H^\lambda | v_1(y) \ra = {1 \over |c(y)|^2} \sum_{i,j=0}^m
\Omega_{ij}^\lambda c_i^* c_j y^{i+j},
\end{equation}
where $\Omega_{ij}^\lambda \equiv \la v_1^i | H^\lambda |v_1^j \ra$
are the ``mixed moments'' of $H$.  Clearly, by operating successively
with H on each $|v_1^i \ra$ $n$ times, the $\Omega_{ij}^\lambda$ can
be evaluated.  This then defines the moments at any $y$,
and thus in principle the exact $L(n,|v_1(y)\ra)$.   This is then
an exact moments Lanczos description of the response surface:
the distribution that results from diagonalizing $L(n,|v_1(y)\ra)$ will
correctly describe the moments $\la v_1(y) | H^\lambda | v_1(y) \ra$,
$\lambda=1,\dots,2n-1$.  We call this the Naive Moments Method (NMM).

The catch is the ``in principle" part: the inversion from moments to
$L(n,|v_1(y)\ra)$ is equivalent to the classical moments problem:
finding a discrete distribution from its moments.
While specific formulae for this
inversion are given in the Appendix, the inversion becomes
increasingly unstable with increasing $n$.  Even with calculations
done in 64-bit precision, the NMM can fail in fewer than 10 iterations.

\textbf{The Legendre Polynomial Moments Method:}  As discussed
in the Appendix, the rapid loss of precision with increasing $n$
in the NMM -- more precisely, in the
inversion to determine the Lanczos matrix and thus
the distribution -- can be traced to the dominance
of the extremum eigenvalues in high-order moments like
$\la H^{2n-1} \ra$.  This suggests reformulating the NMM in 
such a way that the basic physics is preserved -- the simple polynomial 
dependence of moments on $y$ and the use of this dependence
to define moments for all $y$ -- while using combinations of
moments that are better behaved near the extremums.

A possible choice to replace \{$1,H,H^2,....$\} are the Legendre
polynomials in H, \{$P_0(H),P_1(H),P_2(H),...$\}, with the energy
range between the lowest and highest eigenvalues mapped
on to [-1,1], the usual range.  These polynomials have the 
attractive feature that they achieve a fixed magnitude of 1 at the boundaries
of the range.  The specific algorithm we constructed
is described in the Appendix.  While equivalent to the NMM
mathematically, the recurrence relations for determining the 
${\alpha_i,\beta_i}$  from the Legendre polynomial moments 
indeed proved to be more stable.  In some applications 80 iterations
could be performed with little loss of precision.  However, as 
discussed in the Appendix, it sometimes fails more quickly.

Other orthogonal polynomials in $H$ could be used.  An interesting
question is the possibility that some choice might further 
improve the stability.

\textbf{Legendre Coefficients Method:} Rather than carrying through
the procedure of constructing the Lanczos matrix, diagonalizing and
then constructing the strength distribution, it is possible to find
the a truncated expansion of the strength distribution in Legendre
polynomials directly from the starting vector and Hamiltonian.  We
find this method to be stable and effectively reproduce distributions
to great accuracy.  However, the other methods discussed here converge
more rapidly near the ends of the spectrum than in the middle, while the
expansion in Legendre coefficients does not, in general.  The
expansion also lacks positive-definiteness.

In the next section, we discuss a fourth method that does not
attempt to exactly preserve moments, but proves in fact to be
nearly exact (errors less than 0.01\%, typically).  It is stable, positive definite,
and easy to implement.  It is the Lanczos method we recommend for those
needing to generate response surfaces.

\section{The Piecewise Moments Method}
The Piecewise Moments Method (PMM)
is based on the Lanczos matrices and is accurate and
positive-definite, requiring $m+1$ Lanczos calculations
to define the response function over the entire plane.
The basic idea behind the PMM is
to solve the Lanczos problem separately for each of the components
$|v_1^0 \ra, |v_1^1 \ra,\cdots,|v_1^m \ra$ of the vector $|v_1(y) \ra$, while 
incorporating the resolution function directly into the algorithm.
The resolution function provides a prescription for handling 
inner products between Lanczos vectors generated from
different starting vectors.

While the algorithm is general, it is most transparent if formulated
first for a Lorentzian resolution function, making use of the
Lanczos Green's function expansion.  Combining Eqs.~\ref{eqn:res1}
and \ref{eqn:s} yields the Lorentzian-smoothed response function
\begin{equation}
S_L(\omega,q) ={\sigma \over \pi} y^{J-K} e^{-2y} |c(y)|^2 
\sum_{i=1}^N {|\la \psi_{E_i}|v_1(y) \ra|^2 \over (\omega-E_i)^2 + \sigma^2}.
\end{equation}
But this can be rewritten as
\begin{equation}
S_L(\omega,q)={\sigma \over \pi}y^{J-K} e^{-2y}  \la \phi_L(y) | \phi_L(y) \ra
\end{equation}
where
\begin{equation}
\label{eqn:g1}
|\phi_L(y) \ra = {1 \over \omega-H + i \sigma} [c_0|v_1^0 \ra + \cdots c_m y^m |v_1^m \ra ].
\end{equation}
The symmetric form of $\la \phi_L(y) | \phi_L(y) \ra$ guarantees the
positive definiteness of the response function.

At this point we replace the exact Green's function in Eq.~\ref{eqn:g1}
by its Lanczos equivalent, execute the algorithm for each
separate component of $|\phi_L(y) \ra$ as a starting vector, 
and invoke Eq.~\ref{eqn:greens}.
This requires generating the $m+1$ Lanczos matrices $L(n,|v_1^i \ra), i=0, \dots, m$,
then generating the continued fractions from the entries in the tridiagonal matrices,
\begin{eqnarray}
&G(\omega+i\sigma)|v_1^i \ra \rightarrow \hat{G}_n(\omega+i\sigma)|v_1^i \ra&  \nonumber \\
&=\hat{g}_1^i(\omega+i\sigma) |v_1^i \ra  + \cdots +\hat{g}_n^i(\omega+i\sigma) |v_n^i \ra.
\end{eqnarray}
This gives the PMM expression for the response function in the case of a Lorentzian 
resolution function,
\begin{eqnarray}
\label{eqn:PMM}
\hat{S}_L(\omega,q) = {\sigma \over \pi} y^{J-K} e^{-2y} \sum_{i,j=0}^m
c_i^*c_j y^{i+j}  \nonumber \\
\times   \sum_{k,l=1}^n \hat{g}_k^{i*}(\omega+i\sigma) 
\hat{g}_l^j(\omega+i\sigma) \la v_k^i | v_l^j \ra~~~
\end{eqnarray}
Note that Lorentzian PMM does not require a diagonalization of the Lanczos
matrices, as the recurrence relations for the coefficients $\hat{g}_l^j$ use directly
the elements of the Lanczos matrix \cite{Marchisio}.

By construction this result has the right form when $m$=0, i.e., when
the starting vector has a single component.  In this limit $\la v_k^0
| v_l^0 \ra = \delta_{kl}$, and, of course, the procedure would yield
the exact moments.  In more complex cases, the resolution function
plays an important role in interpreting the overlap of Lanczos
vectors, as we discuss below.

First, however, it is helpful to generalize the PMM for other resolution functions.
We use the Gaussian of Eq.~\ref{eqn:res1} as an example, as other cases are similar:
\begin{equation}
S_G(\omega,q) = {1 \over \sigma \sqrt{2 \pi}} y^{J-K} e^{-2y} \la \phi_G(y) | \phi_G(y) \ra
\end{equation}
where
\begin{equation}
|\phi_G(y) \ra =  e^{-(\omega-H)^2/4\sigma}  [c_0|v_1^0 \ra + \cdots c_m y^m |v_1^m \ra ].
\label{eqn:Gaussexp}
\end{equation}

Now one applies the Lanczos algorithm, evaluating $L(n,|v_1^i\ra), i=0,\dots,m$.  These
$m+1$ $n$-dimensional matrices are then diagonalized, yielding the Lanczos 
eigenvalues \{$\hat{E}_l^i$\} and \{$\hat{\psi}_{\hat{E}_l}^i$\}, $l=1,\dots,n$.   The appropriate
complete set can be inserted in each term of Eq.~\ref{eqn:Gaussexp}, e.g.,
\begin{eqnarray}
\sum_{l=1}^n e^{-(\omega-H)^2/4\sigma} |\hat{\psi}_{\hat{E}_l}^i\ra \la \hat{\psi}_{\hat{E}_l}^i|v_1^i\ra \nonumber \\
=\sum_{l=1}^n e^{-(\omega-\hat{E}_l^i)^2/4\sigma} |\hat{\psi}_{\hat{E}_l}^i\ra \hat{\psi}_{\hat{E}_l}^{i*}(1).~
\end{eqnarray}
One thus derives the PMM result
\begin{eqnarray}
\label{eqn:PMMgen}
\hat{S}_G(\omega,q) = {1 \over \sigma \sqrt{2 \pi}} y^{J-K} e^{-2y} \sum_{i,j=0}^m
c_i^*c_j y^{i+j}~~~~~~~ \nonumber \\
\times \sum_{k,l=1}^n e^{-[(\omega-\hat{E}_k^i)^2 + (\omega-\hat{E}_l^j)^2]/4 \sigma}
\hat{\psi}_{\hat{E}_k}^i(1) \hat{\psi}_{\hat{E}_l}^{j*}(1)  \la \hat{\psi}_{\hat{E}_k}^i |
 \hat{\psi}_{\hat{E}_l}^j \ra~~~
\end{eqnarray}
This ``general form'' of the PMM requires diagonalization of the Lanczos matrices, just
as in Eq.~\ref{eqn:strengthdistribution}.  Any other resolution function can be substituted 
for the Gaussian: the prescription is to ``take the square root'' of the resolution function,
letting it act symmetrically left and right, thereby preserving the positive-definiteness of the
response function.

An exercise helpful in understanding this result is to consider the
limit $n \rightarrow N$.  In this limit the Lanczos eigenvectors will
converge to true eigenvectors of $H$, regardless of the starting
vectors.  The scalar products in Eq.~\ref{eqn:PMMgen} then reflect the
resulting orthonormality, independent of the indices $i$ and $j$.  Now
consider $n$ very close, but not equal, to $N$.  In this case one
expects, for $i \not= j$, to find nearly identical eigenvectors
$|\hat{\psi}_k^i \ra$ and $|\hat{\psi}_k^j \ra$ with nearly equal but
still distinct eigenvalues.  If $\sigma$ in the resolution function of
Eq.~\ref{eqn:strengthdistribution} is significantly larger than the
eigenvalue splitting, the differences will not matter: the overlap
will be evaluated just as if we had continued to the limit $n
\rightarrow N$.  The response function will not change as additional
iterations are done.  Conversely, if $\sigma$ is smaller, the response
function clearly will continue to evolve, as additional iterations are
done, until those eigenvalues do become degenerate, on a scale defined
by $\sigma$.

In general, when $n \ll N$, the situation will be more complicated,
with a number of states within some energy range overlapping between
Lanczos calculations done with different starting vectors, $i \not=
j$.  But the Lanczos algorithm does properly capture the strength for
each starting vector, omitting high-frequency information.  By
introducing a resolution function directly into the algorithm we make
high-frequency information irrelevant, provided $n$ is large enough
for the desired $\sigma$.  Thus, even though the high-frequency
information omitted for $i \not= j$ may be somewhat different for
these two starting vectors, nevertheless one would expect the
procedure described above to converge.  This expectation can be tested
numerically.

\section{Numerical Tests of the  Piecewise Moments Method}
To do so we have picked the test case of electromagnetic form factors for $^{28}$Si, 
evaluated in a full $sd$-shell calculation using the Brown-Wildenthal interaction \cite{brown}.  The
operator notation follows in part Ref.~\cite{walecka}.

The Coulomb response function is defined as
\begin{eqnarray}
\label{eqn:coulomb}
C_J(\omega,q) = \sum_f  \delta(\omega-E_f) \times~~~~~ \nonumber \\
| \la f; J || \sum_{i=1}^{A} M_J(q\vec{r}_i) ({1 +\tau_3(i) \over 2}) || g.s.; 0 \ra |^2
\end{eqnarray}
where $||$ denotes a reduced matrix element and the sum extends over a
complete set of $sd$-shell final states, $| f; J \ra$, of the
requisite angular momentum $J$.  The operator $M_J^M$ is
\begin{equation}
\label{eqn:mop}
M_J^M(q \vec{r}) = j_J(qr) Y_{JM}(\Omega_r)
\end{equation}
Because the ground state has $J=0$, it is particularly
simple to rewrite this in the form of Eq.~\ref{eqn:responsefunction}
\begin{eqnarray}
C_J(\omega,q)=\sum_f \delta(\omega-E_f) \times~~~~~~~~~ \nonumber \\
|\la f; J M=0| [J] \sum_{i=1}^A M_{J0}(q\vec{r}_i) ({1 +\tau_3(i) \over 2}) |g.s.; 00 \ra|^2\nonumber \\
\equiv \sum_f \delta(\omega-E_f) |\la f;JM=0|O(q)|g.s.;00\ra|^2~~~
\end{eqnarray}
where $[J]=\sqrt{2J+1}$.  The operator $O(q)$, introduced to make the analogy with
Eq.~\ref{eqn:responsefunction} clear, can be evaluated using the 
tables of Donnelly and Haxton \cite{donnelly}.  Its second-quantized form is
\begin{eqnarray}
\label{eqn:o1}
{[J] \over \sqrt{4 \pi}} y^{(J-2)/2} e^{-y} \sum_{\alpha \beta}
(-1)^{j_\alpha-m_\alpha} \left(\begin{array}{ccc}
j_\alpha & J & j_\beta \\ -m_\alpha & 0 & m_\beta \\\end{array} \right) \times \nonumber \\
p_{M_J}^{\alpha \beta}(y) a^{\dagger}_{\alpha_p} a_{\beta_p} ~~~~~~~~~~~~~~~~~~~~~~~
\end{eqnarray}
where $p_{M_J}^{\alpha \beta}(y)$ is the polynomial tabulated in the tables.  The sums
over single-particle states $\alpha$ and $\beta$ are restricted to protons
because of the isospin projection $(1 +\tau_3(i))/2$ and to the $sd$-shell, due to the
nuclear model.  In the case of $C0$, the inert core nucleons must also be included.
As the maximum single-particle angular momentum in the $sd$-shell is $j$=5/2,
$J$=0,2, and 4 Coulomb multipoles are possible.

Similar operators can be obtained for the transverse electric response function
$E_J(\omega,q)$
\begin{eqnarray}
\label{eqn:o2}
{[J] \over \sqrt{4 \pi}} {q \over M} y^{(J-2)/2} e^{-y} \sum_{\alpha \beta}
(-1)^{j_\alpha-m_\alpha} \left(\begin{array}{ccc}
j_\alpha & J & j_\beta \\ -m_\alpha & 0 & m_\beta \\\end{array} \right) \times \nonumber \\
\left[ (p_{\Delta'_J} ^{\alpha \beta}(y) + {\mu_p \over 2} p_{\Sigma_J}^{\alpha \beta}(y))
{a^\dagger}_{\alpha_p} a_{\beta_p} + {\mu_n \over 2} p_{\Sigma_J}^{\alpha \beta}(y)
{a^\dagger}_{\alpha_n} a_{\beta_n} \right]~~~~
\end{eqnarray}
and transverse magnetic response function $M_J(\omega,q)$
\begin{eqnarray}
\label{eqn:o3}
{[J] \over \sqrt{4 \pi}} {q \over M} y^{(J-1)/2} e^{-y} \sum_{\alpha \beta}
(-1)^{j_\alpha-m_\alpha} \left(\begin{array}{ccc}
j_\alpha & J & j_\beta \\ -m_\alpha & 0 & m_\beta \\\end{array} \right) \times \nonumber \\
\left[ (p_{\Delta_J} ^{\alpha \beta}(y) - {\mu_p \over 2} p_{\Sigma'_J}^{\alpha \beta}(y))
{a^\dagger}_{\alpha_p} a_{\beta_p} - {\mu_n \over 2} p_{\Sigma'_J}^{\alpha \beta}(y)
{a^\dagger}_{\alpha_n} a_{\beta_n} \right].~~~~
\end{eqnarray}
In these equations $\mu_p$ and $\mu_n$ are the proton and neutron magnetic
moments.  Normally the charge and magnetic single-nucleon couplings are
described by form factors, but we will treat all couplings as fixed (point nucleon
limit), as the purpose of this study is the modeling of the nuclear momentum 
dependence.  If these couplings are given a common momentum dependence,
that factor could be added to the nuclear results we present below.

In these equations the underlying single-particle operators are
\begin{eqnarray}
\Delta_J^M(q \vec{r}) &=& \vec{M}_{JJ}^M(q \vec{r}) \cdot {1 \over q} \vec{\bigtriangledown} \nonumber \\
{\Delta'}_J^M(q \vec{r}) &=& -i \left[ {1 \over q} \vec{\bigtriangledown} \times \vec{M}_{JJ}^M(q \vec{r})
\right] \cdot {1 \over q} \vec{\bigtriangledown} \nonumber \\
\Sigma_J^M(q \vec{r}) &=& \vec{M}_{JJ}^M(q \vec{r}) \cdot \sigma \nonumber \\
{\Sigma'}_J^M(q \vec{r}) &=& -i \left[ {1 \over q} \vec{\bigtriangledown} \times \vec{M}_{JJ}^M(q \vec{r}) \right] \cdot \sigma
\end{eqnarray}
where the spherical Bessel vector harmonic operator is
\begin{equation}
\vec{M}_{JL}^M(q \vec{r}) =  j_L(qr) \vec{Y}_{JL1}^M(\Omega_r).
\end{equation}

The polynomials $p^{\alpha \beta}$ can again be found in the tables.
The overall polynomial behavior -- that is, the y-dependence other
than the overall multiplicative factors shown explicitly in
Eqs.~\ref{eqn:o1},\ref{eqn:o2},\ref{eqn:o3} -- are as follows
\begin{eqnarray}
C0&:& (y^1,y^2,y^3) \nonumber \\
C2/E2&:& (y^1,y^2) \nonumber \\
C4/E4&:& y^1 \nonumber \\
M1&:& (y^0,y^1,y^2) \nonumber \\
M3&:& (y^0,y^1) \nonumber \\
M5&:& y^0
\end{eqnarray}
The most complex cases, the C0 and M1 response functions,
have three contributing terms in $y$.  Thus a maximum of three
Lanczos calculations is needed to define any
electromagnetic response function over the $(\omega,q)$ plane,
for PMM calculations in the $sd$-shell.

\begin{figure}
\includegraphics[width=9.4cm]{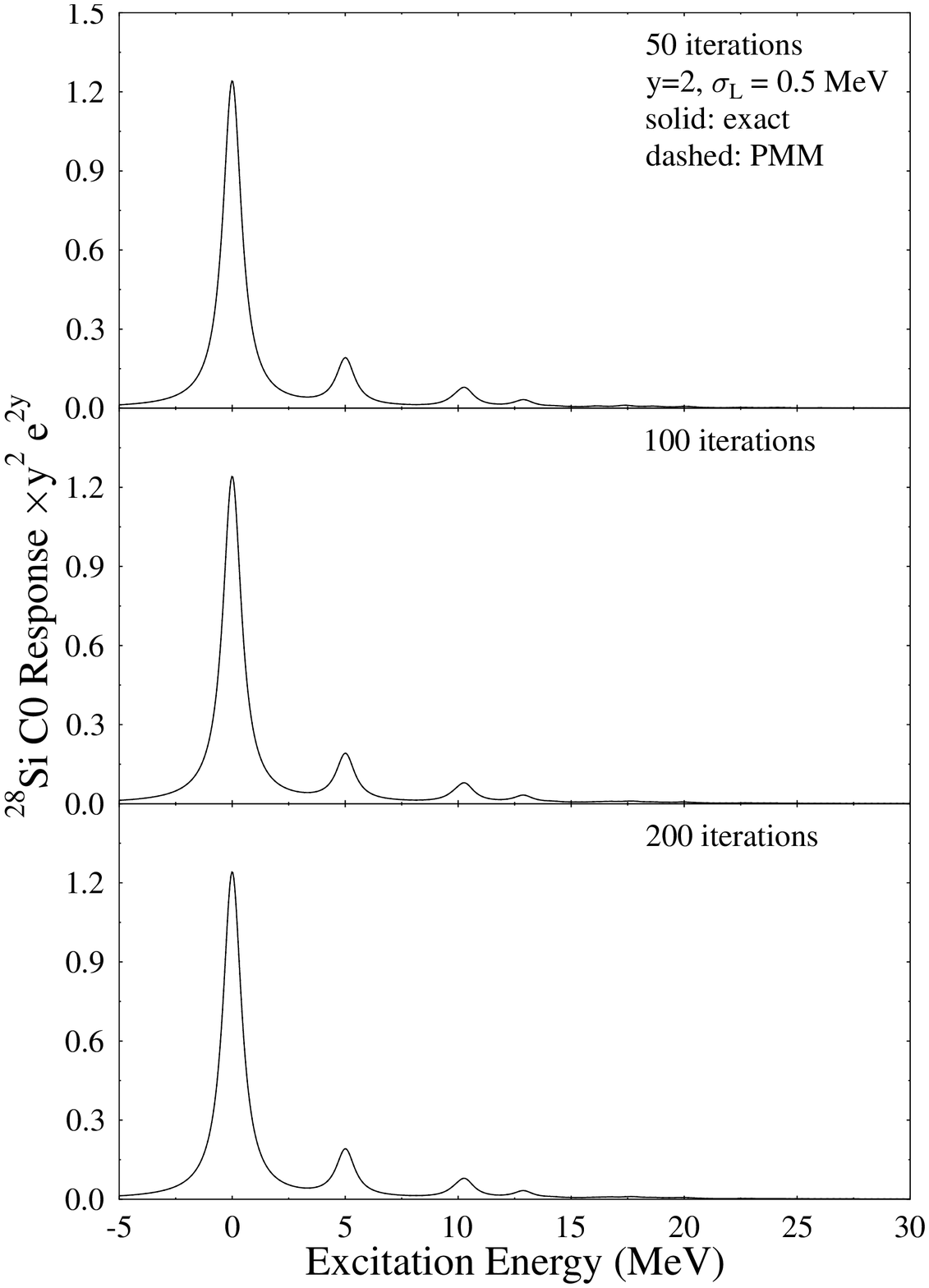}
\includegraphics[width=9.4cm]{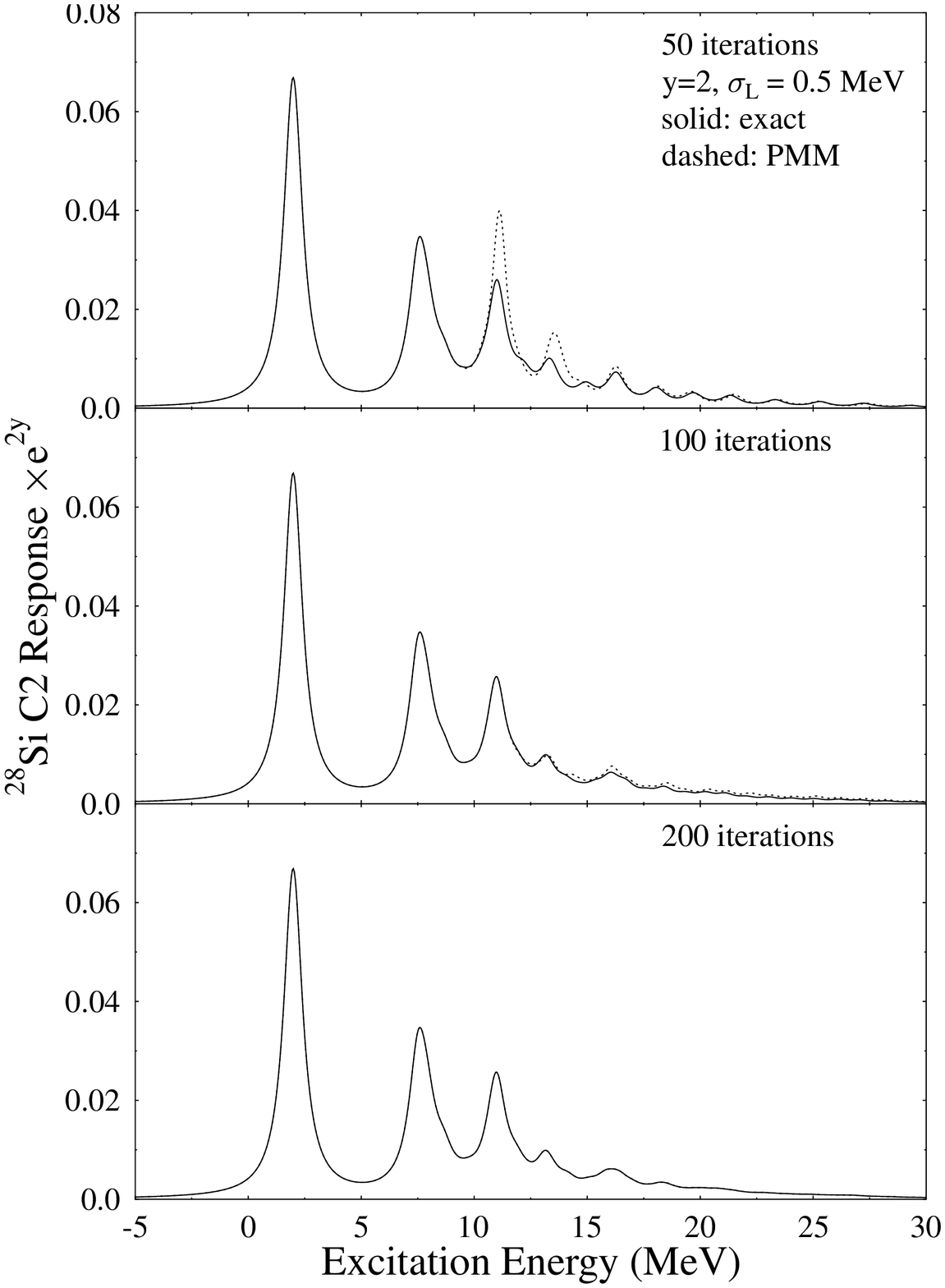}
\caption{Comparison of an exact moments calculation (solid line)
with the PMM (dashed) method for the C0 and C2 response functions
along the $y$=2 line in the response plane.}
\label{fig:C_1}
\end{figure}

\begin{figure}
\includegraphics[width=9.4cm]{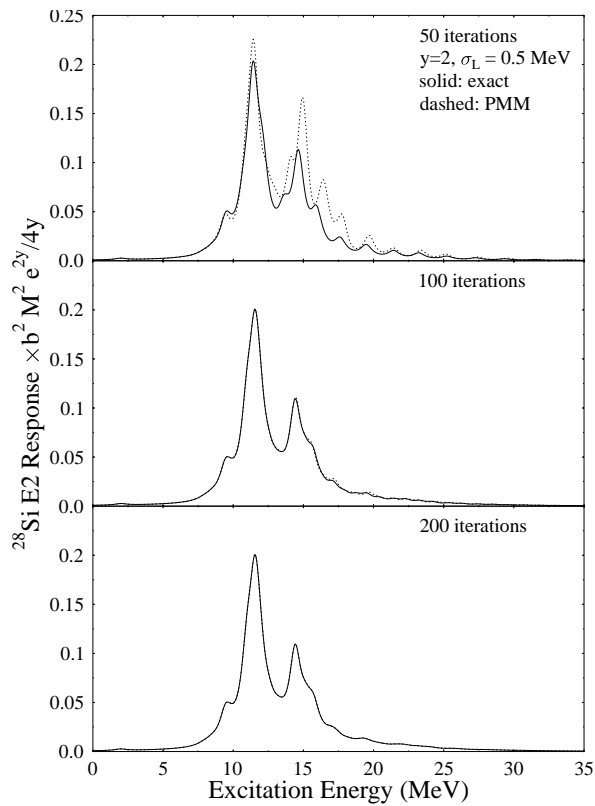}
\caption{As in Fig. 1, only for the E2 response.}
\label{fig:E_2}
\end{figure}

\begin{figure}
\includegraphics[width=9.4cm]{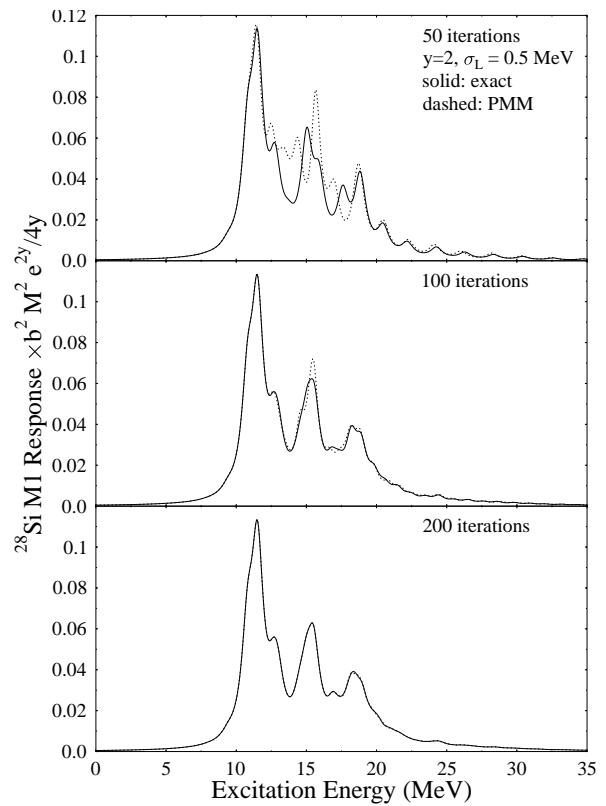}
\includegraphics[width=9.4cm]{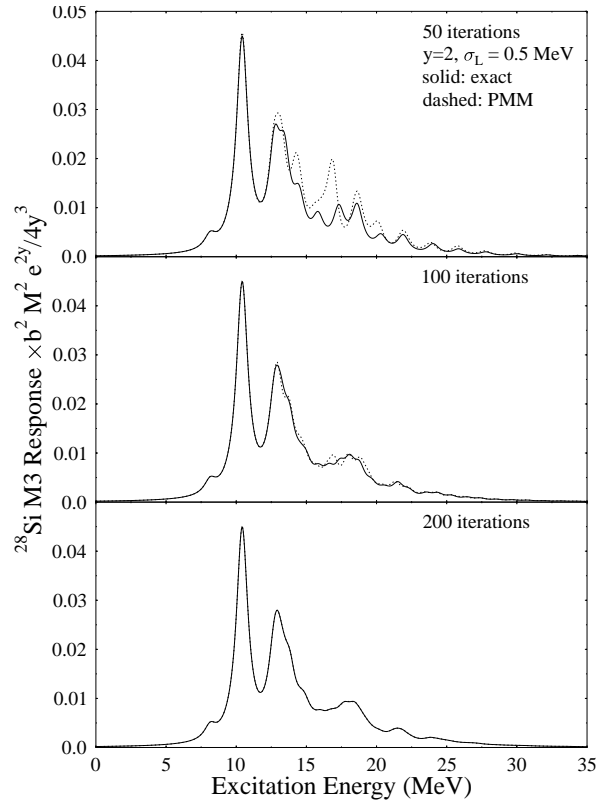}
\caption{As in Fig. 1, only for the m1 and M3 responses.}
\label{fig:M_3}
\end{figure}

\begin{figure}
\includegraphics[width=9.4cm]{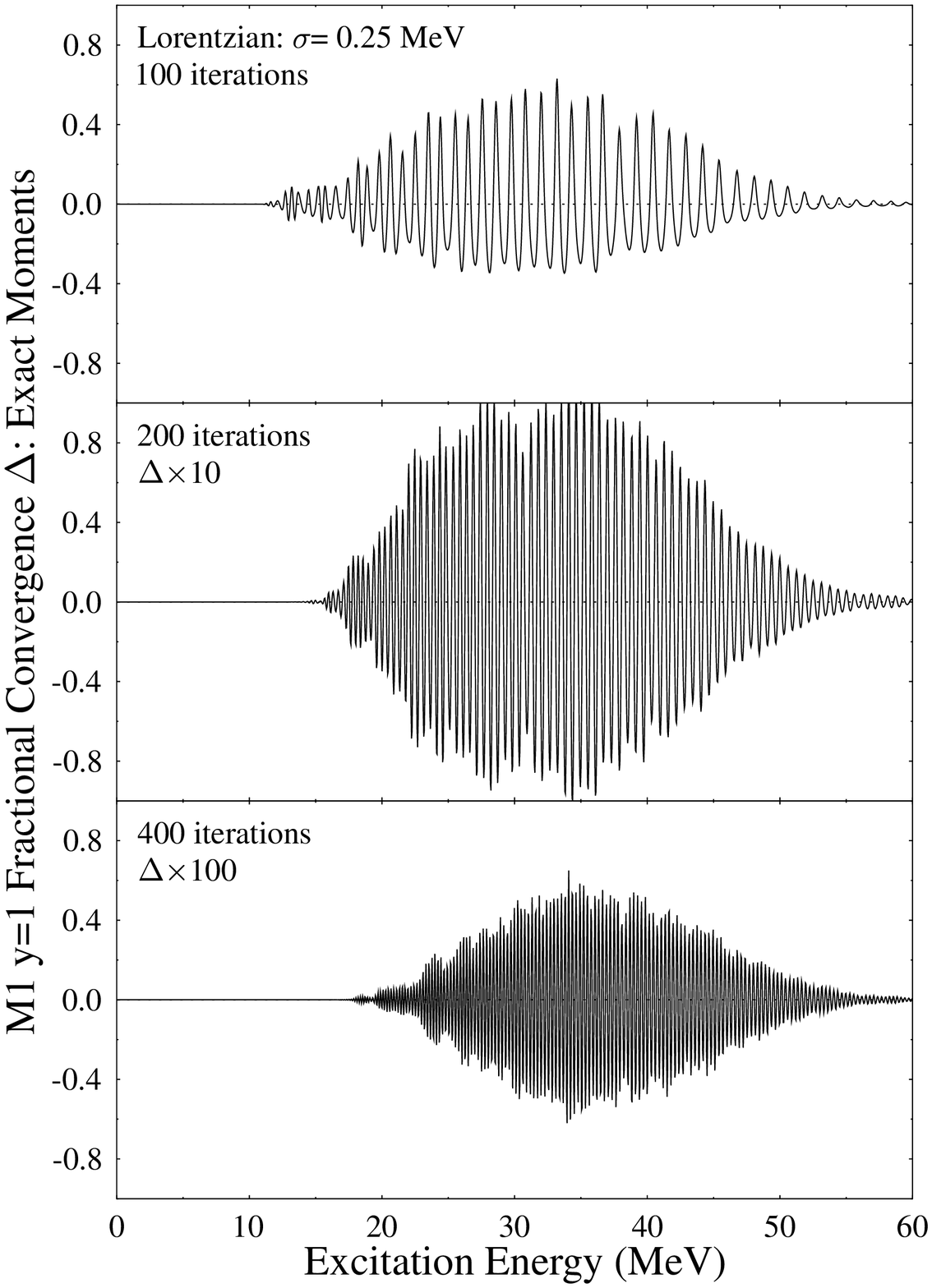}
\includegraphics[width=9.4cm]{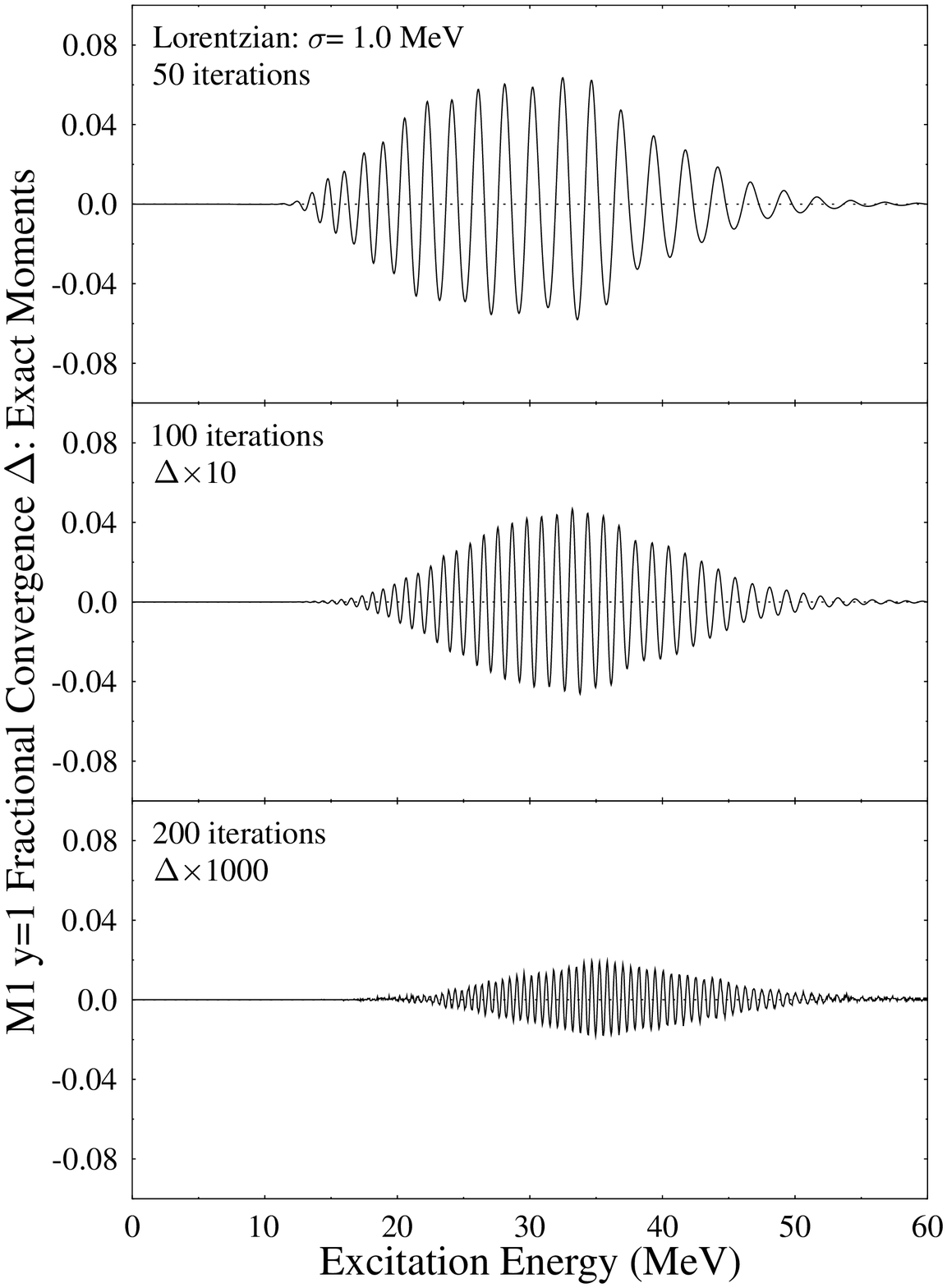}
\caption{Comparison of the convergence of the M1
response for $y$=1 for  $\sigma$ = 0.25 and 1.0 MeV in an exact moments calculation.}
\label{fig:M1exact_4}
\end{figure}

\begin{figure}
\includegraphics[width=9.4cm]{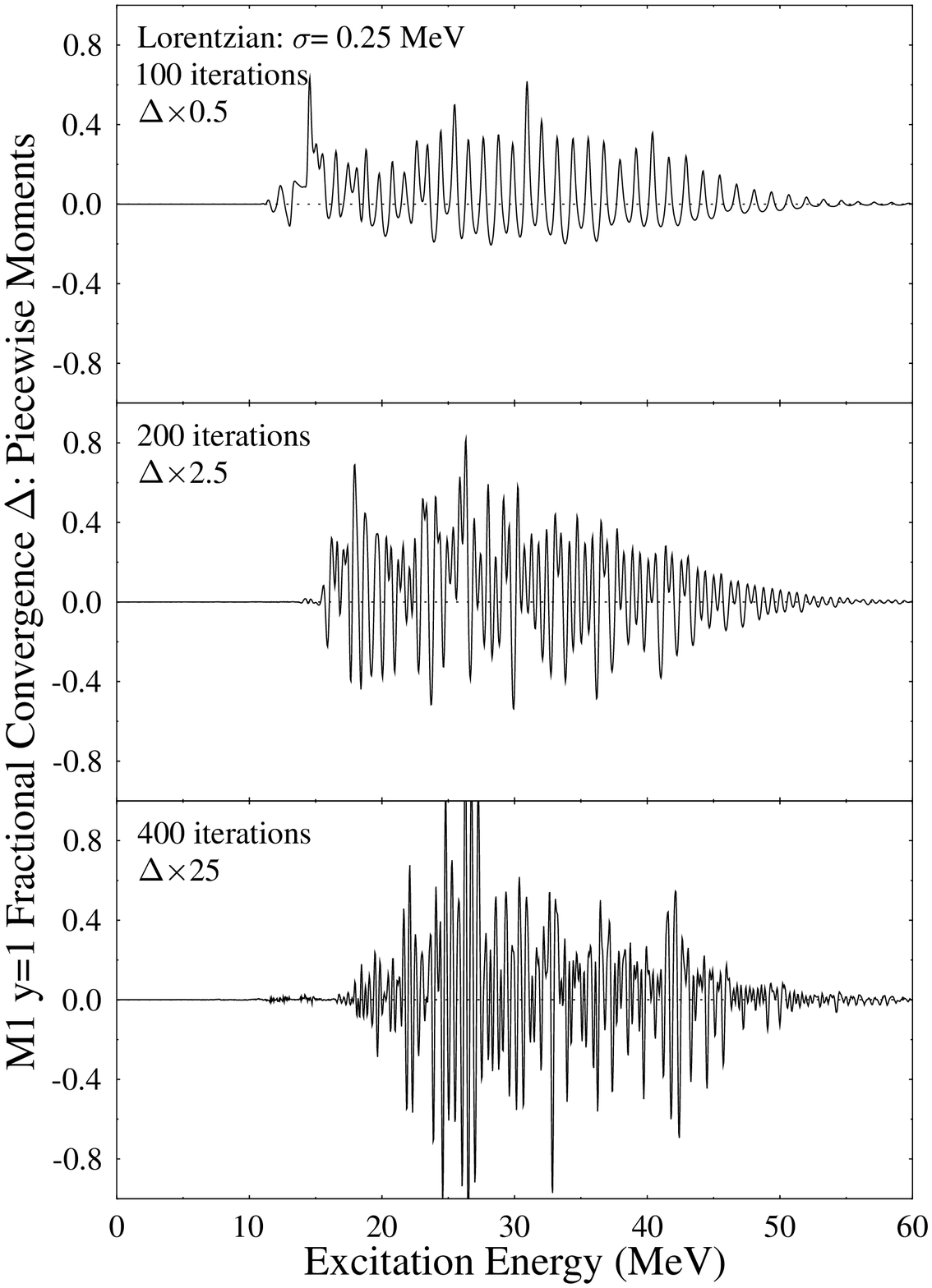}
\includegraphics[width=9.4cm]{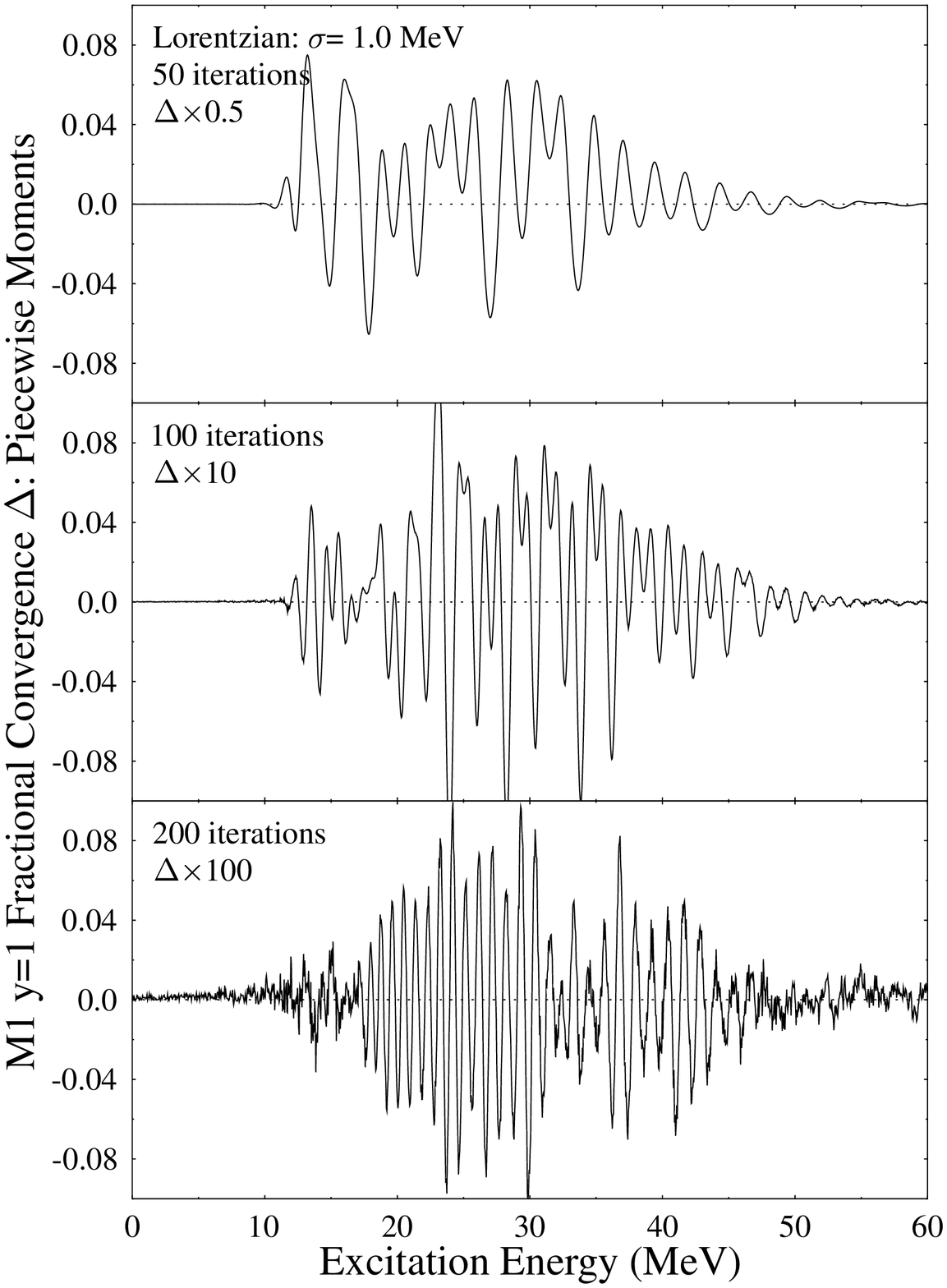}
\caption{As in Fig. 4, only for the PMM calculation.}
\label{fig:M1_5}
\end{figure}

\begin{figure}
\includegraphics[width=9.4cm]{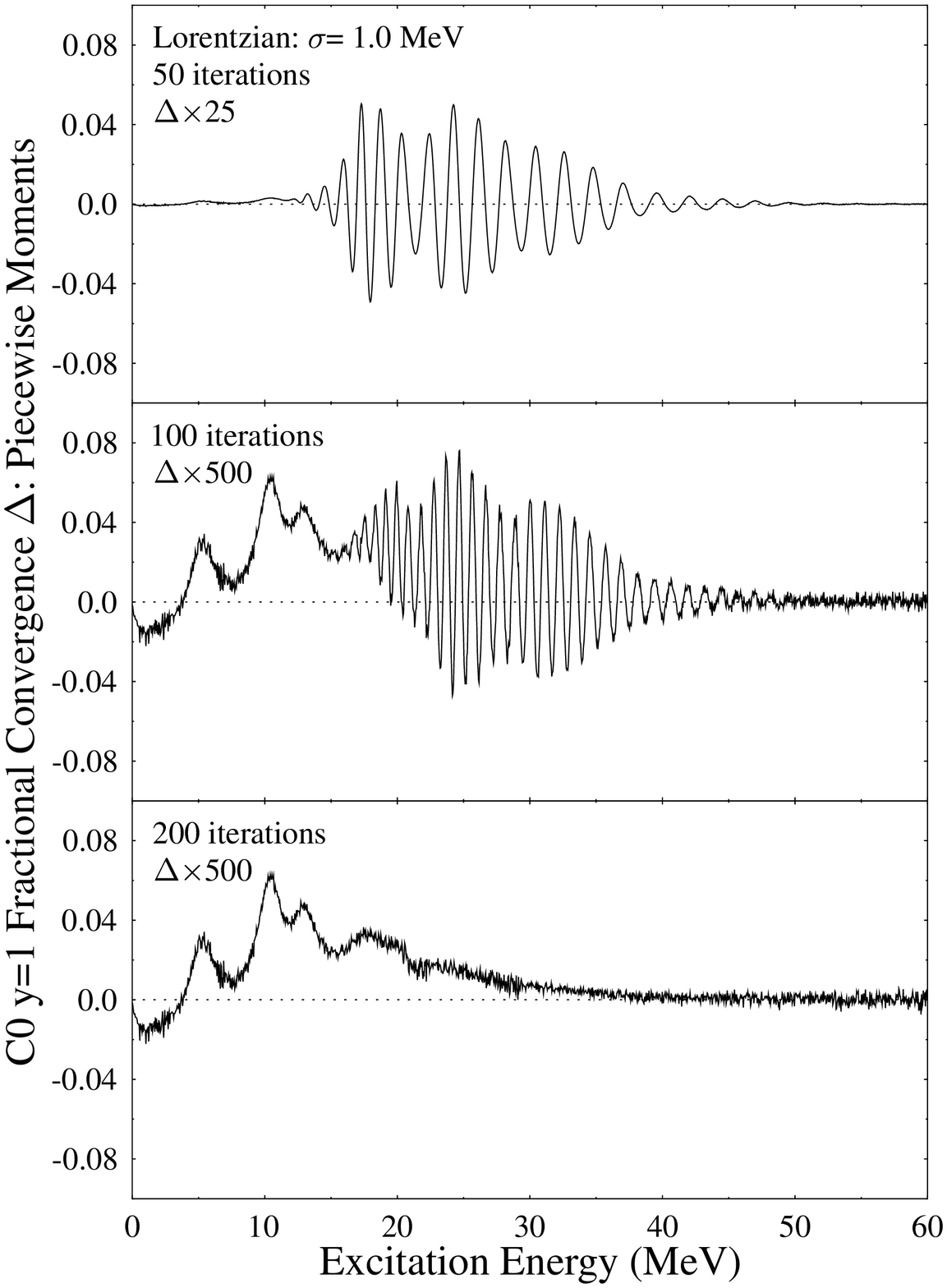}
\caption{Convergence of the PMM C0 response for $y$=1 and
$\sigma$=1.0 MeV.  A small difference persists
near the dominant extremum eigenvalues.}
\label{fig:C0_6}
\end{figure}

All of these response functions were evaluated with the PMM,
for several resolution-function widths ($\sigma$=1.0, 0.5, and 0.25 MeV)
and for $n$ ranging from 50 to 400 iterations.  The results we show
all assume a Lorentzian for the resolution function.
The accuracy of the PMM was first tested by examining cuts 
corresponding to $y$=constant in the $(\omega,q)$ plane.
For such trajectories, $O(y)$ is fixed, so that Eq.~\ref{eqn:responsefunction}
can be used -- an exact moments treatment.   This will allow us to
test our qualitative argument that the PMM should be numerically difficult to
distinguish from an exact moments treatment, provided $n$ is large enough
for the given $\sigma$.  Figures 1-3 show the results for the
C0, C2, E2, M1, and M3 response functions -- the cases with more than
one component in the vector $|v_1(y)\ra$ -- for $y$=2 and
$\sigma$=0.5 MeV, and with $n$=50,100, and 200.

In those cases where there is significant structure in the response
function, discrepancies are apparent at $n$=50, but these disappear as
more iterations are added.  In all cases, by $n$=200 differences
between an exact moments calculation and the PMM result are not
readily discernible on the scale of the graphs.  Note that the
differences at $n$=50 reflect that fact that {\it neither} the exact
moments calculations nor the PMM results are fully converged.  Similar
results were obtained for $y$=1.  This kind of test could be made by
anyone using the PMM, to guarantee, for the chosen resolution function
and $\sigma$, that a sufficient number of iterations have been done to
produce the desired accuracy.

The next series of figures provide a more detailed look at the
convergence and its dependence on $\sigma$.  Fig. 4
provides benchmarks for exact moments calculations, the residual discrepancies between
calculations for $n$=50, 100, 200, and 400 iterations
and one with $n$=600, which we will take as a fully
converged result.  The figures show the convergence in the
case of the M1 response evaluated along the line $y$=1
at the resolutions $\sigma$ = 0.25 and 1.0 MeV.
The behavior is just as one would expect.  The missing contributions
are oscillatory, with a frequency that is roughly proportional to
$n$ but virtually independent of $\sigma$, for the range we explored.
The envelope of the oscillations decreases 
with increasing $n$, shrinking by more than an order of 
magnitude for every additional 50 iterations for $\sigma$=1.0 MeV.
The decrease slows to about a factor of 5 every 100 iterations
for the more taxing calculation with $\sigma$=0.25 MeV.

Similar calculations, not shown, were  performed for the
C0 response, which has much less structure than the M1 
response.  The results are qualitatively similar to that shown
in Fig. 4, though the convergence of the envelope
is a factor of $\sim$ 30 more rapid. 

Figure 5, the analog of Fig.~4, gives the M1
PMM residuals (again relative to an exact moments calculation
with $n$=600).  While again an oscillatory pattern emerges with
a frequency like that of Fig. 4, its structure is less
regular.  This is the result of the
interference between the three Lanczos patterns
that contribute to the PMM result, corresponding to the starting vectors $|v_1^0\ra$, $|v_1^1\ra$, and
$|v_1^2\ra$.  The PMM envelopes tend to be a factor of two
to three more extended than those from the exact moments
calculation, though in one case the difference is larger.

Figure 6, the residuals for the PMM calculation for
the C0 response, with $y$=1 and $\sigma$ = 1.0 MeV,
show an interesting effect.  Through most
of the spectrum the same oscillatory features and diminishing
envelope with increasing $n$ are seen.  But in the low-energy
region, a persistent feature has converged by $n$=100.
It appears dominantly positive, but as the fractional convergence
is graphed and as this response is dominated by the ground
state, where the differential is negative, this is misleading.
The low-energy C0 response is
characterized by isolated extremum eigenvalues, as is apparent
from Fig. 1, with the ground state carrying the entire response from
$|v_1^0\ra$ ($y \rightarrow$ 0).  These appear to be conditions
that allow for a small deviation of the PMM from the results
of an exact moments calculation, in converged calculations.
However, the deviation is small, less than 0.01\%.  Very similar
effects were found in the C0 responses for $\sigma$ = 0.5 MeV
and 0.25 MeV, with the discrepancies reaching 0.03\% and
0.06\% in these cases, respectively.

Finally, we show a series of PMM results for the response surface.
Figure 7 shows contour plots for the C0, C2, and C4 responses,
while Figs. 8 and 9 give the three-dimensional
projections of the E2 and E4 and the M1, M3, and M5 responses, respectively.
The general shift of strength to larger $y$ with increasing 
multipolarity is apparent.  In most applications of the PMM, such
response surfaces, determined as a function not only of $(\omega,q)$
but also of the oscillation parameter $b$, would be the end result
of the calculations.

\begin{figure}
\includegraphics[width=8.4cm]{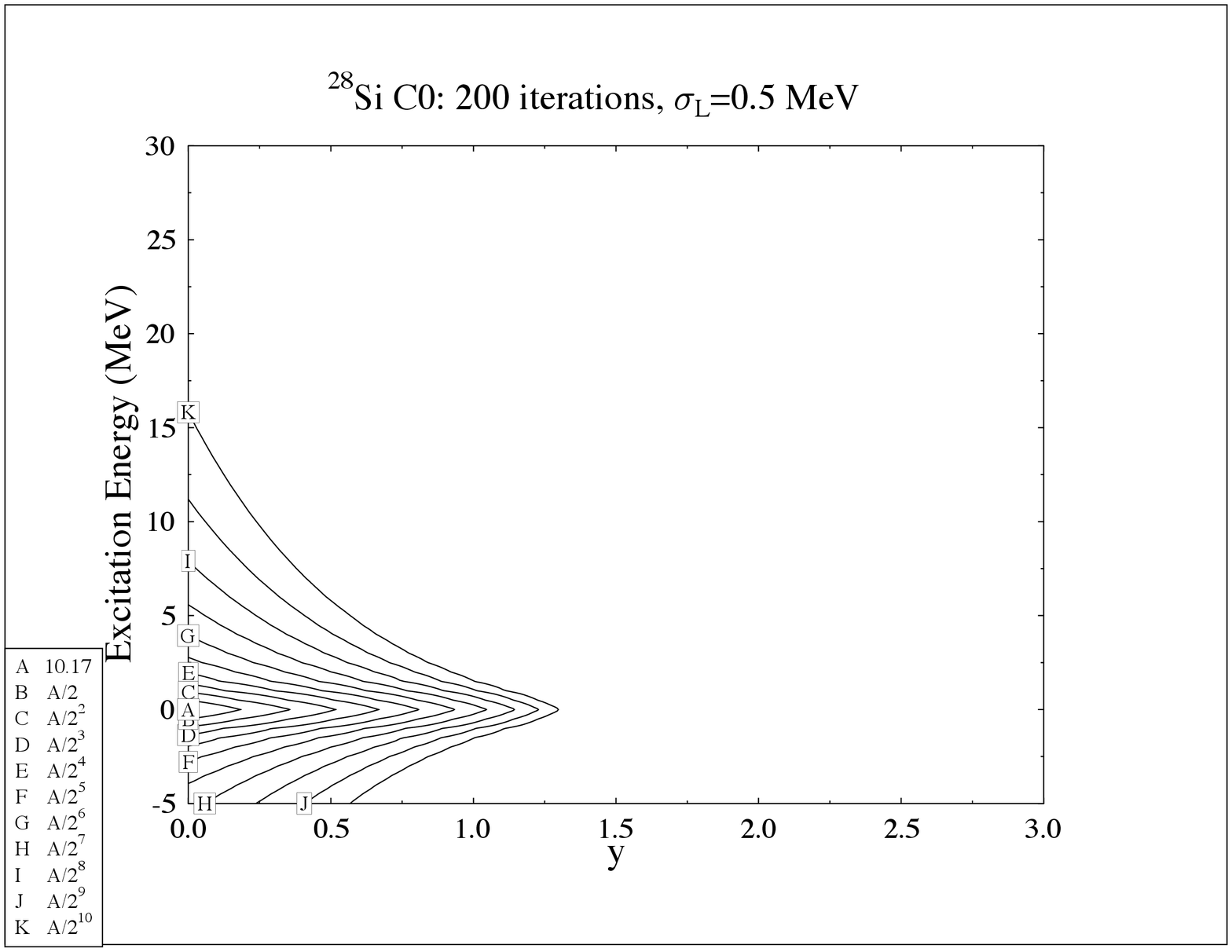}
\includegraphics[width=8.4cm]{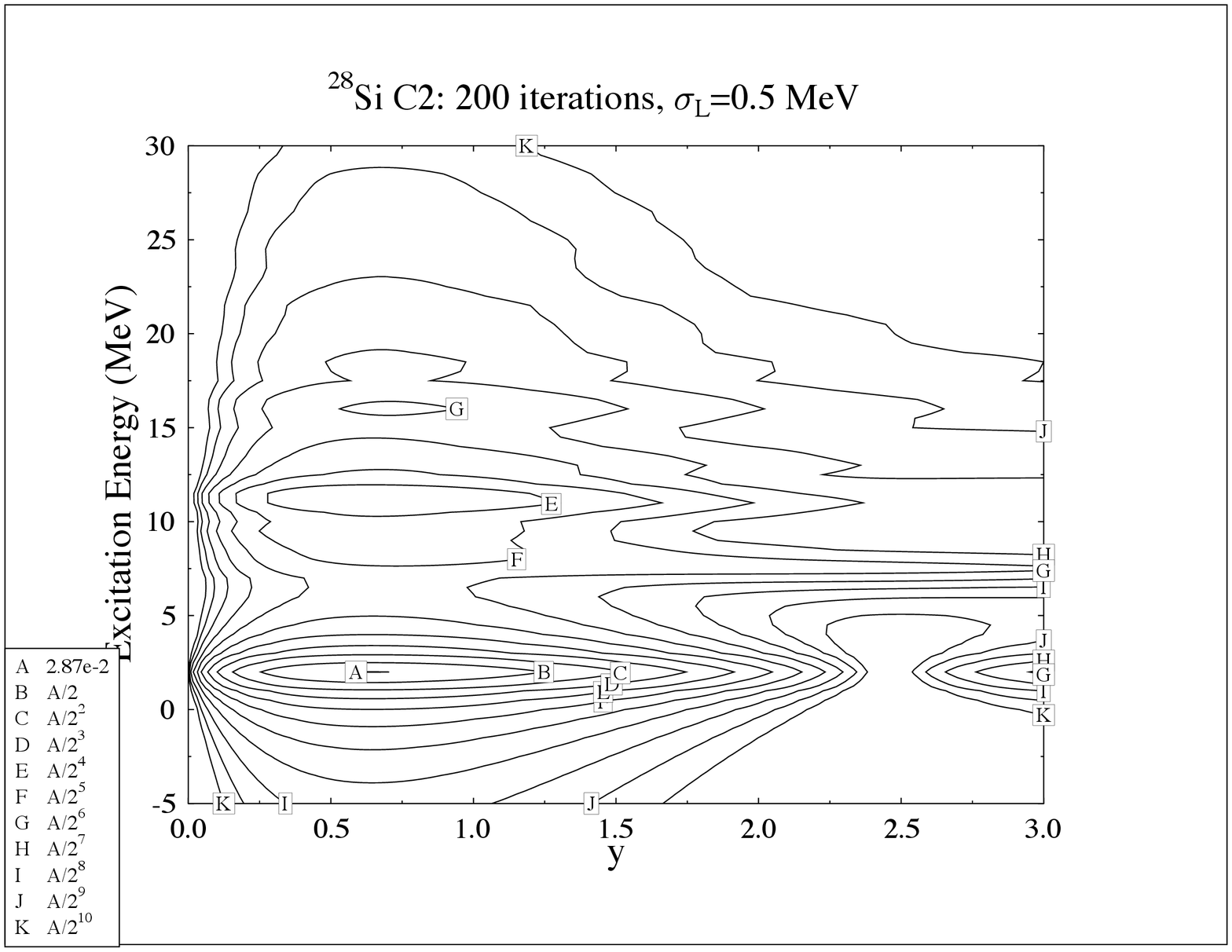}
\includegraphics[width=8.4cm]{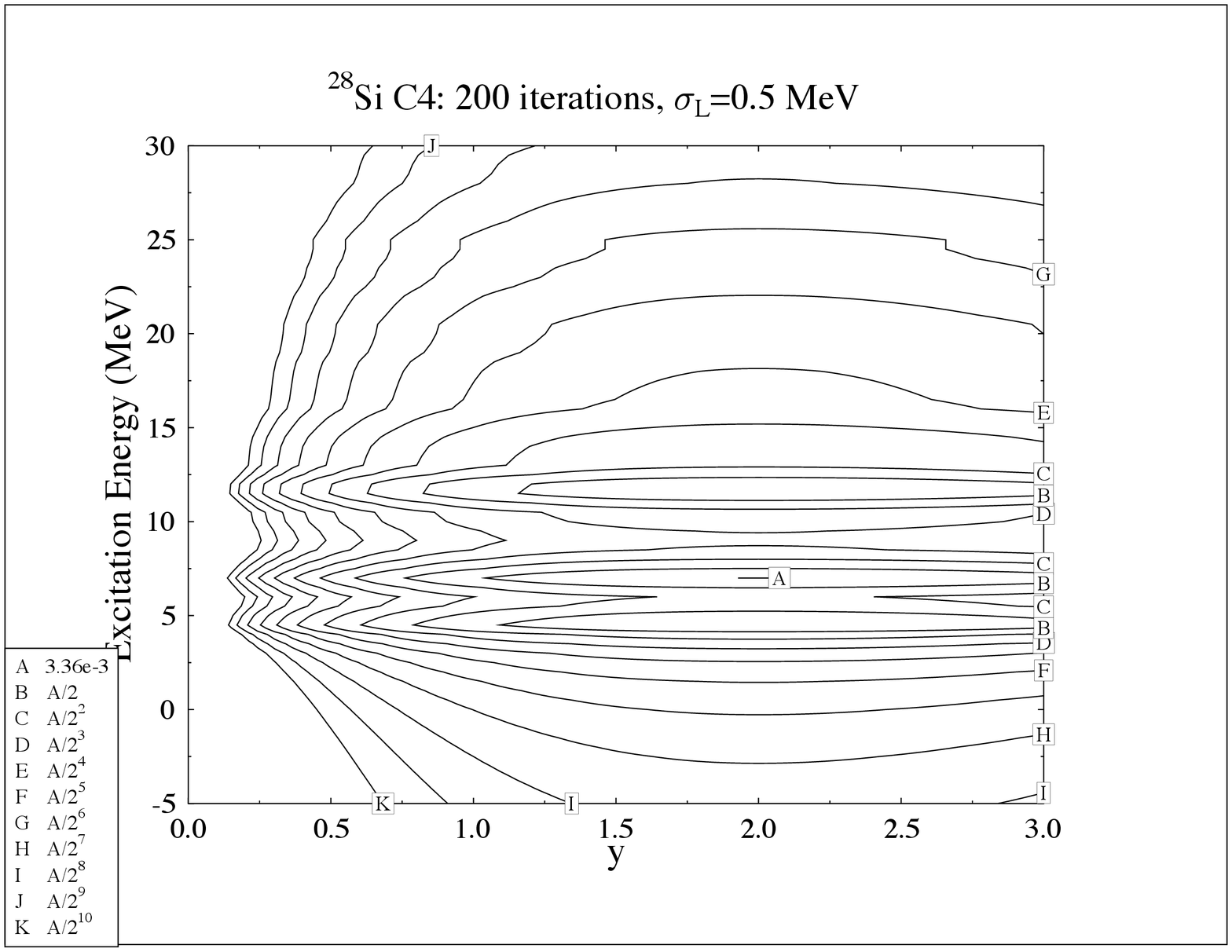}
\caption{The PMM response surfaces for the C0, C2, and
C4 multipoles, with contours drawn at successive factors of 0.5 
of the maximum, until 0.001 is reached.}
\label{fig:C_contour_7}
\end{figure}

\begin{figure}
\includegraphics[width=9.5cm]{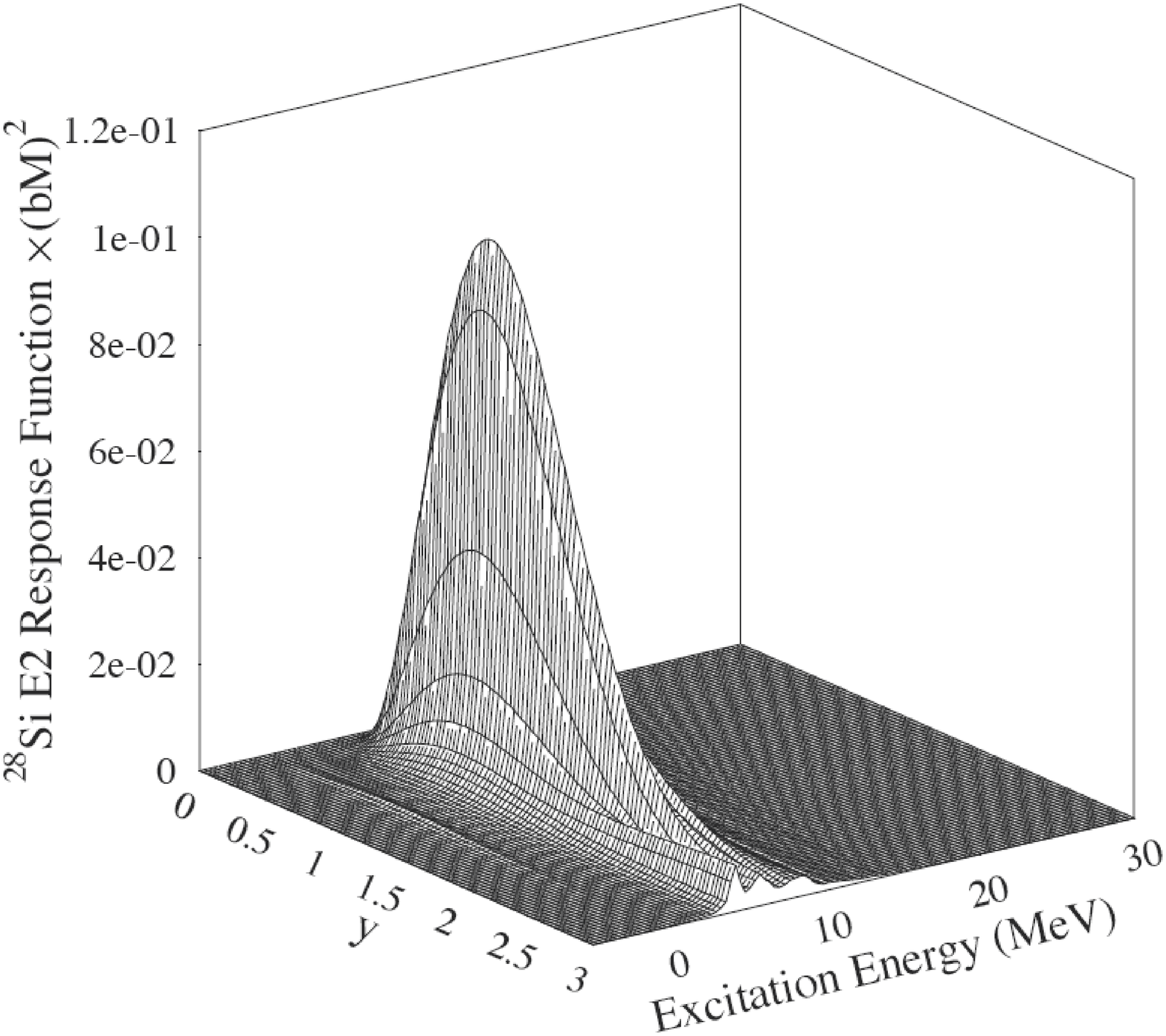}
\includegraphics[width=9.5cm]{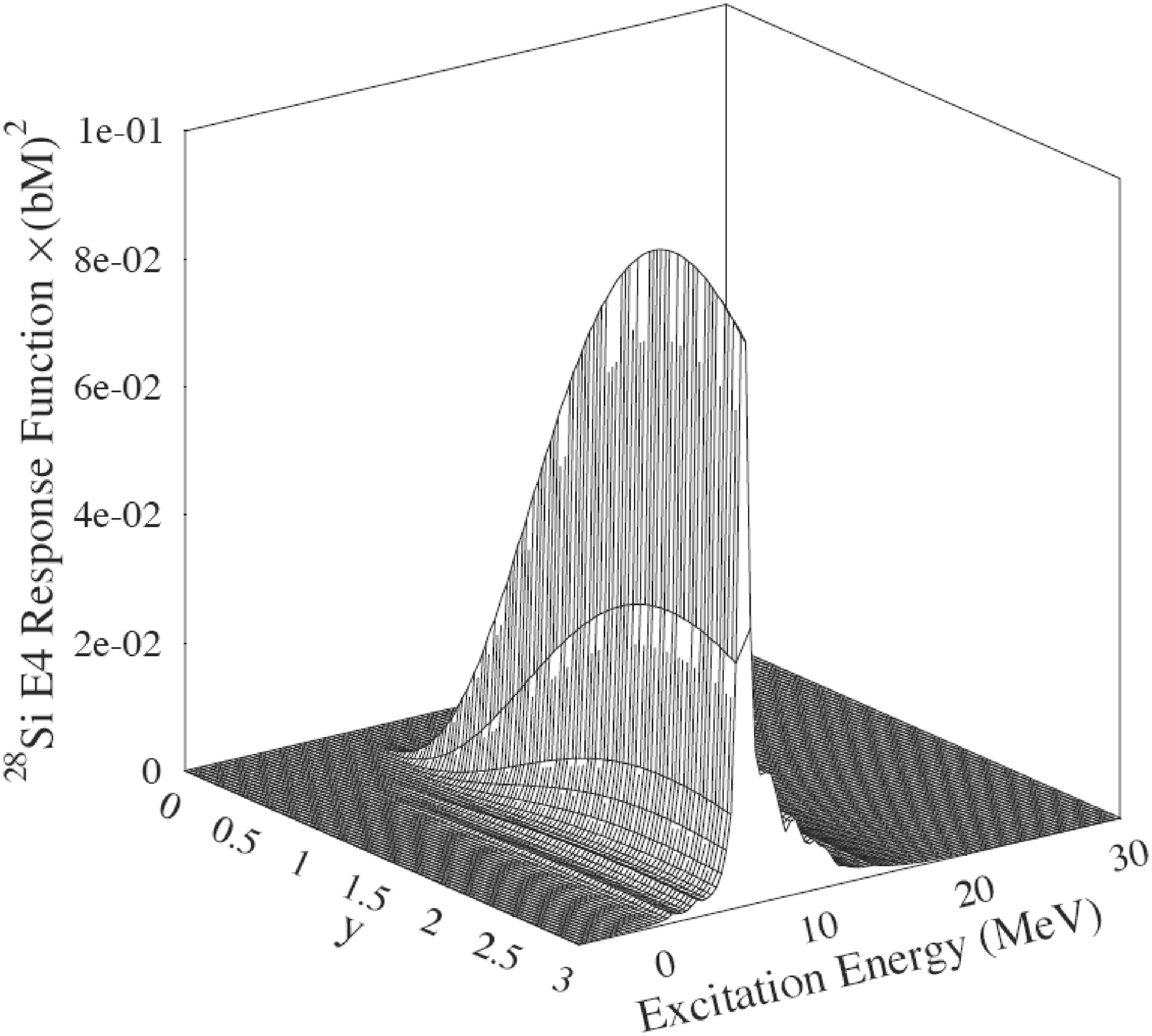}
\caption{The PMM response surface for the E2 and
E4 multipoles.}
\label{fig:E_3d_8}
\end{figure}

\begin{figure}
\includegraphics[width=9.5cm]{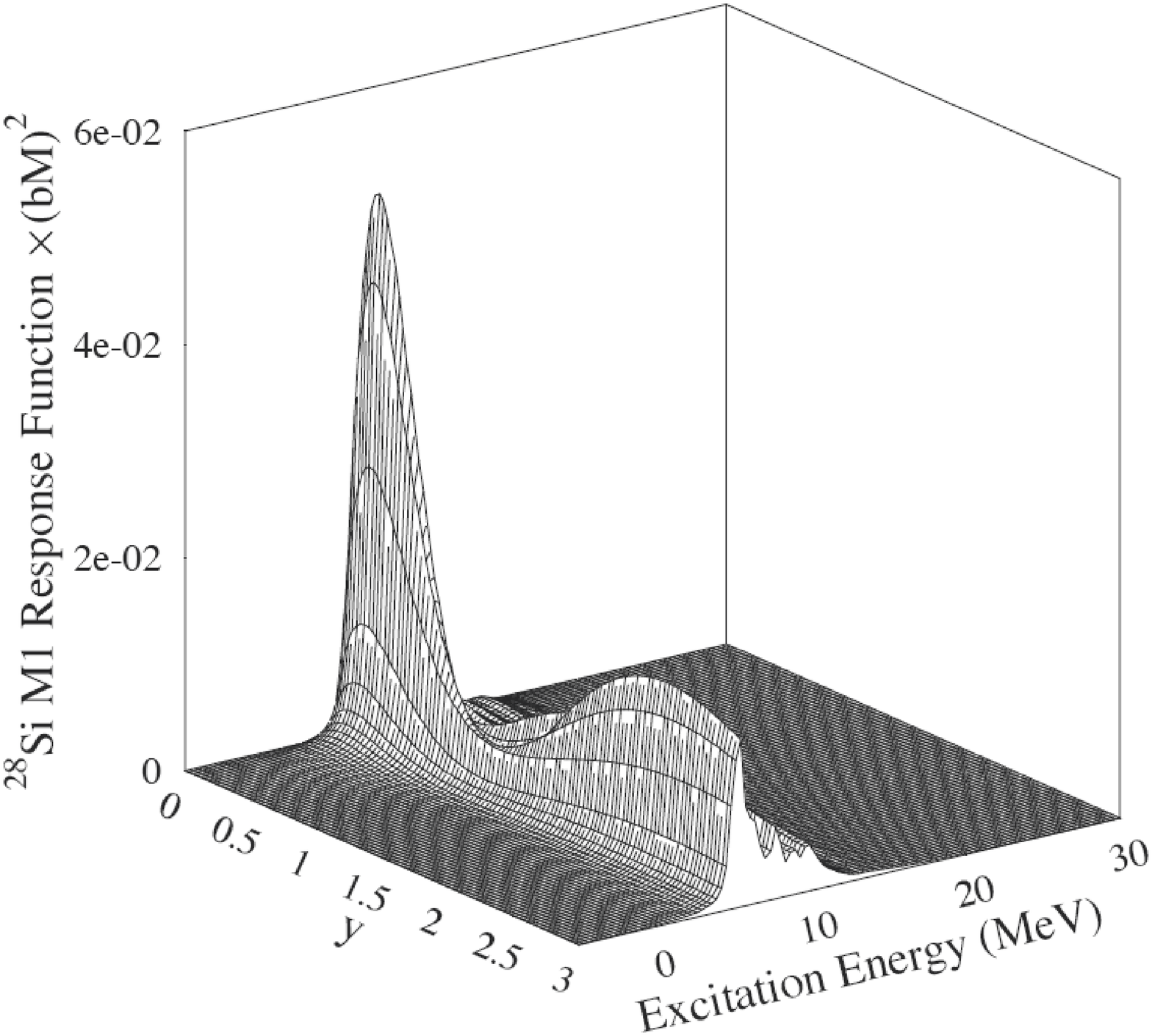}
\includegraphics[width=9.5cm]{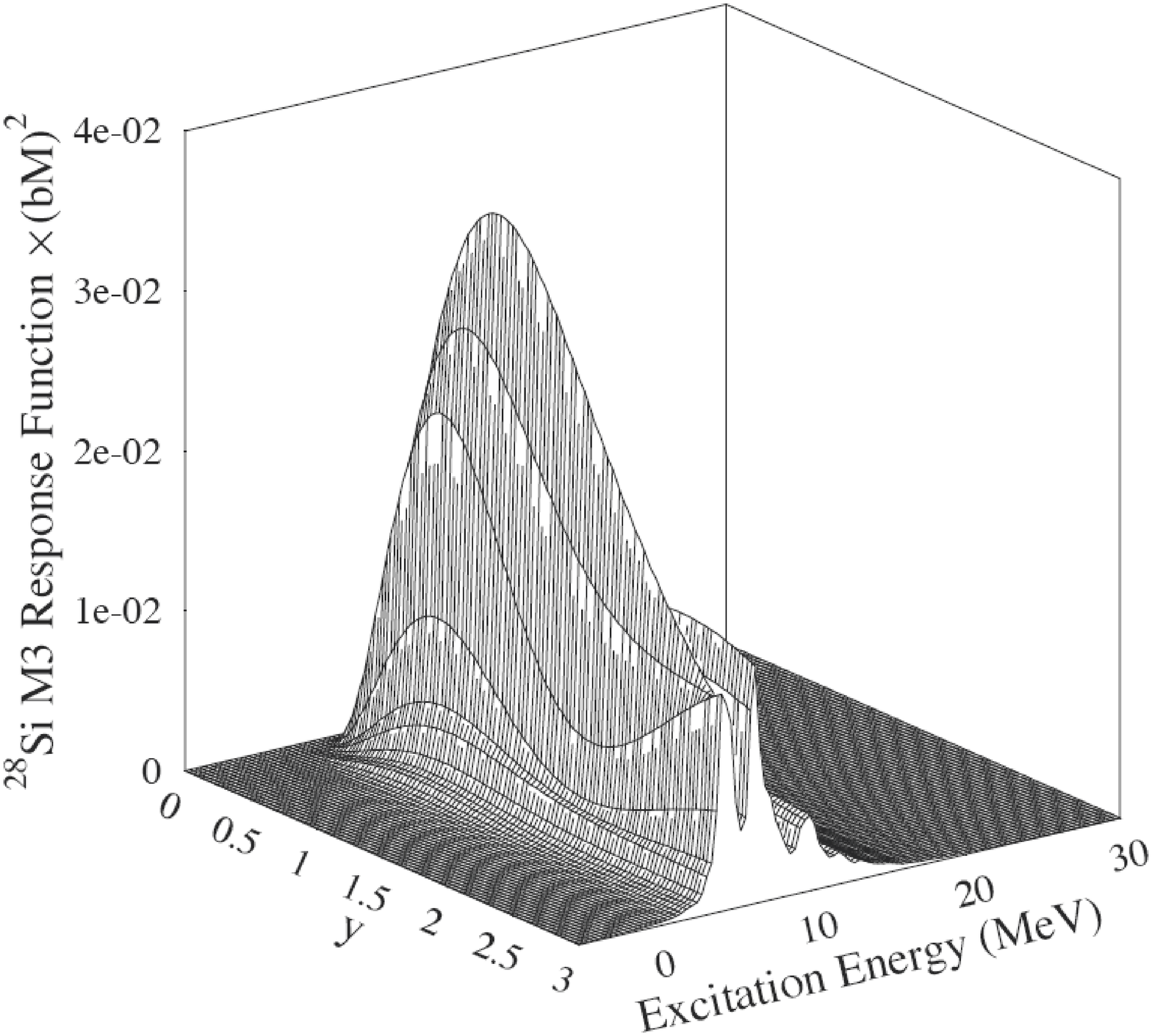}
\includegraphics[width=9.5cm]{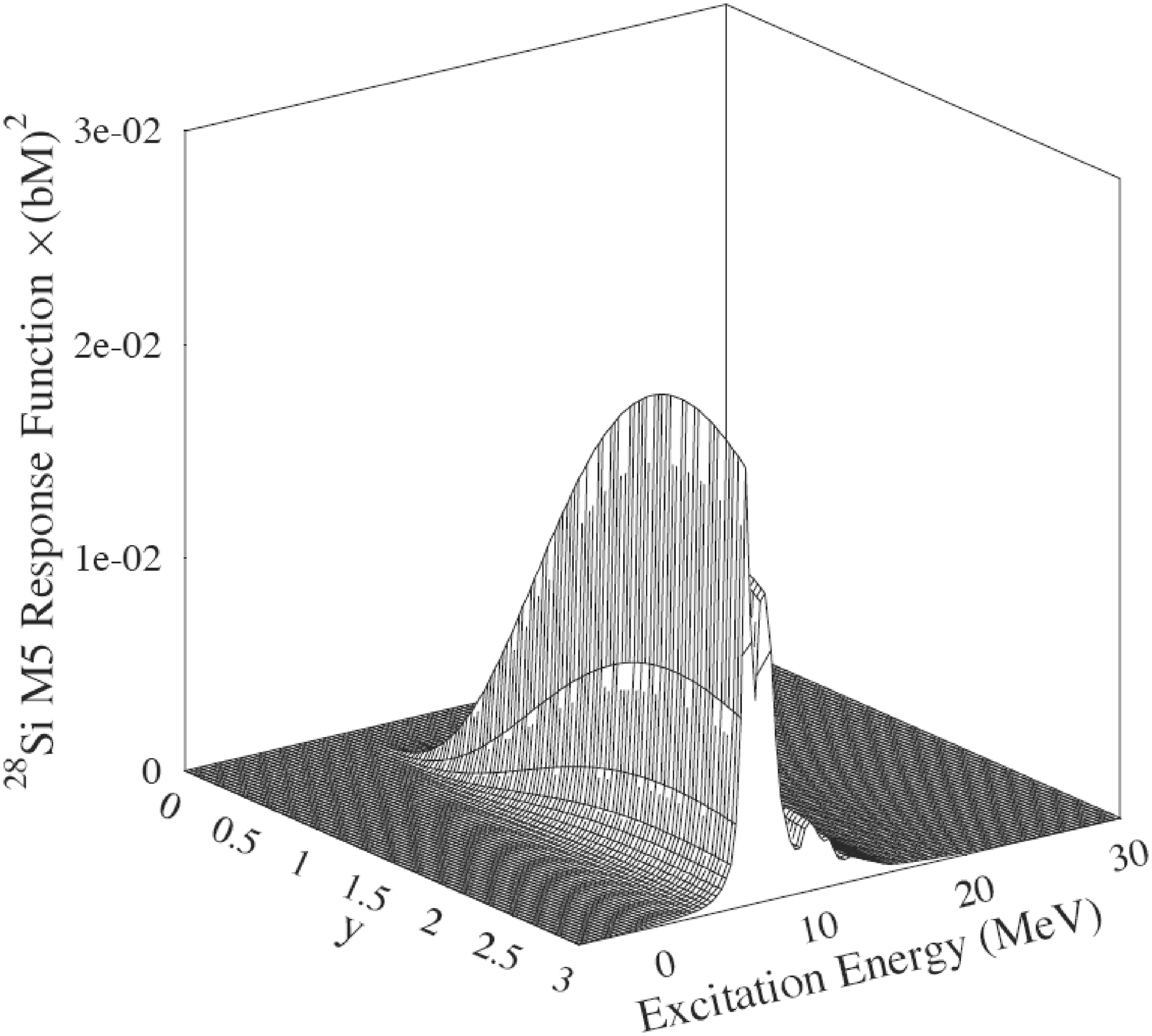}
\caption{As in Fig. 8, only for the M1, M3, and M5 responses.}
\label{fig:M_3d_9}
\end{figure}

\section{Conclusions}
An attractive property of the matrix elements of electroweak single-particle
operators between harmonic oscillator states is that they can be evaluated
analytically, yielding a simple form that includes a finite polynomial in
$y=(qb/2)^2$.  This property has been exploited frequently in calculations
of discrete transition amplitudes.  In this paper we have shown that
moments methods exploiting this property can very efficiently characterize
the entire response surface over the $(\omega,q)$ plane.  This can be 
viewed as an important extension of well-known moments techniques
for determining the distribution of Gamow-Teller strength along the
line $q=0$ in the $(\omega,q)$ plane.

While the method was motivated by the observation that moments 
must have a polynomial form in $y$, algorithms we designed to 
exactly preserve the moments were found to be numerically unstable,
if a sufficient number of iterations $n$ were done.
This difficulty is the well-known ``classical moments problem,"
the determination of a discrete distribution from knowledge of the
distribution's moments.  An alternative method, called the
Piecewise Moments Method, was introduced in
which each polynomial component of the starting vector was treated
separately in the Lanczos procedure, 
while also incorporating a resolution function directly into
the algorithm.   The method is extremely stable, positive definite,
and effectively equivalent, numerically, to an exact moments 
reconstruction.  After convergence -- which typically requires $\sim$ 200
iterations for a resolution $\sigma \sim$ 0.5 MeV -- we found maximal differences 
between the PMM and exact moments calculations of about 0.01\%,
in one case (the C0 multipole).  The number of iterations $n$ 
required for convergence increases with decreasing $\sigma$.

We noted that the absence of a moments method to reconstruct 
the response in the full $(q,\omega)$ plane had previously led to the
use of very simple nuclear models, so that state-by-state sums
over transitions could be performed.  Clearly the PMM will now
allow theorists to perform analogous calculations using state-of-the-art
shell-model wave functions in very large model spaces.  
The efficiency with which the PMM constructs the response
function over the response plane is
impressive.  In the example we explored -- the electromagnetic form
factors for $^{28}$Si in the $sd$-shell -- the most complicated
multipoles, C0 and M1, required only three Lanczos steps.

Another aspect of the method, discussed in Section II, is that it can be
viewed as a numerical effective theory in the sense that exactly that
information needed to reconstruct $S(\omega,q)$ can be systematically extracted
from exceedingly complicated nuclear structure calculations.
Specifically, if $n$ is the number of iterations and $m$ the rank of $p(y)$,
one needs $(m+1)(2n-1)$ tridiagonal Lanczos matrix elements and
$[m(m-1)/2]n^2$ Lanczos vector overlaps to implement Eq.~\ref{eqn:PMM},
for example.  Once this information is in hand, a simple routine could
be coded to generate $S(\omega,q)$ as a continuous function of $\omega$
and $q$, as well as of the oscillator parameter $b$ and $\sigma$.  (The one
caveat in that for any fixed $n$, there will be some minimum $\sigma$ 
beyond which the number of iterations $n$ performed would not be sufficient.)
In effect, $S(\omega,q)$ would be no more complex, in numerical
calculations, than some analytically known leptonic scattering kernel.
One application we have in mind is neutrino reactions in core-collapse
supernova modeling.  This approach would allow one to model such
reactions with state-of-the-art shell model techniques, yet produce a 
result sufficiently simple that it could be used on-line within a 
sophisticated supernova code.  The fact that $S(\omega,q)$ is given
as a function of $\omega$ and $q$, rather than as a grid of values, is
also important in such applications.  This will allow supernova modelers
to more easily guarantee properties such as detailed balance in
reactions and inverse reactions -- which if not enforced exactly, can
lead to energy generation and other spurious physics when weak interactions
are in equilibrium.  Detailed balance can be more difficult to enforce
in cases where $S(\omega,q)$ is provided on some discrete numerical grid.

We would like to mention three follow-up studies that we think will
make the present work more valuable.  First is the extension to weak
interactions.  This is relatively simple,  as only two new operators
(in addition to those we have treated in Eqs.~\ref{eqn:mop} and \ref{eqn:o3})
arise when axial currents are added (in the standard nonrelativistic treatment where
charges and currents are kept to order $1/M$) \cite{walecka,donnelly}.  One of the main motivations
of the present paper is to develop a technique that can be applied to
neutrino response functions in the energy range up to $\sim$ 1 GeV.

A second question is the loss of unitarity for momentum-dependent 
operators acting in finite shell model spaces.  This issue does not arise
for the standard application to the Gamow-Teller operator $g_A^{eff} \sigma \tau_\pm$,
for any complete $n \hbar \omega$ shell-model space, as the operator
has no cross-shell matrix elements.  But for more general operators, as the momentum
increases, so does the strength of excitations to states outside the model space.
For a vector $O(y) |g.s. \ra$ it is simple exercise to calculate the loss
of probability due to excitations outside the model space.  If one fails to
take into account such loss of probability, trends in inclusive cross sections
as a function of $q$ will be distorted.  Thus it is important to either correct
for such effects, or to evaluate their size to estimate uncertainties in results.

A third issue is the overcompleteness of shell-model spaces due to
spurious center-of-mass motion.  This can be troublesome when dealing
with one-body operators that can excite center-of-mass excitations.
While this issue did not arise in our example of $^{28}$Si because we
confined ourselves to the 0 $\hbar \omega$ $sd$-shell space, it will
in more complicated spaces.  If the space is separable -- e.g., any $n
\hbar \omega$ shell-model calculation with oscillator wave functions
-- the standard technique, when dealing with discrete calculations, is
to remove spurious states by adding to the Hamiltonian a term $\alpha
H_{CM}$, where $H_{CM}$ is the center of mass Hamiltonian and $\alpha
\sim$ 100 a coefficient chosen to ``blow out" spurious states from the
low-energy spectrum \cite{whit2}.  This works well for converged
Lanczos calculations: the addition of such a term forces low-lying
eigenvalues to have the center-of-mass in the $1s$ state.  One would
need to explore numerically whether a similar technique might allow
some approximate separation of spurious excitations over the full
spectrum: we are not aware of any studies of this method apart from
the case of converged extremal eigenvalues.  Alternatively, if the
shell-model Hamiltonian is (properly) translationally invariant and
$|g.s.\ra$ has been constructed so that its center of mass is in the
$1s$ state, center-of-mass excitations in the vector $O(y)|g.s.\ra$
could be removed at the outset.

\acknowledgments

This work was supported by the U.S. Department of Energy, Office of
Nuclear Physics, under contract W-31-109-ENG-38 (KMN) and
grants DE-FG02-00ER-41132 and DE-FC02-01ER-41187 (SciDAC)
(WCH and KZ).

\appendix*
\section{Alternative Methods}
In this Appendix we discuss in more detail alternative Lanczos response 
function methods that we have explored.  This discussion might
be useful to those that would like to further explore some of the
stability issues we encountered.  The first two approaches
have as their goal a reconstruction of $S(\omega,q)$ that exactly
captures the information in the $2n-1$ moments.  The main drawback
in both methods is instability for moderate $n$, connected with the 
classical moments problem (the inversion from moments to a distribution).
A third method is discussed which deals directly with distributions 
while also preserving a specified number of moments. 

\subsection{Naive Moments Method}
As discussed in the main body of the paper, the NMM
begins with a starting vector of the form
\begin{equation}
\label{eqn:again}
c(y)|v_1(y)\ra = c_0|v_1^0\ra + c_1 y |v_1^1\ra + \dots + c_m y^m |v_1^m\ra
\end{equation}
from which the moments $\la v_1(y) | H^\lambda | v_1(y) \ra$, $\lambda=1,\dots,2n-1$, can be
determined for any $y$ once the mixed moments $\Omega_{ij}^\lambda$
have been evaluated.  As knowledge of the moments as a function of $y$
is mathematically equivalent to knowledge of $L(n,|v_1(y)\ra)$, the response
function can thus be evaluated as a function of both $y$ and $\omega$.

On first sight, it appears that the inversion from moments to the
tridiagonal matrix can be easily done.  It has been shown \cite{akhiezer} that the
elements of the Lanczos matrix are related to the moments by
determinants so that
\begin{equation}
\label{eqn:alpha}
\alpha_i=\mathcal{M}_{i-1}/\mathcal{L}_{i-1}
-\mathcal{M}_{i-2}/\mathcal{L}_{i-2}
\end{equation}
and
\begin{equation}
\label{eqn:beta}
\beta_i=\mathcal{L}_i^{1/2}\mathcal{L}_{i-2}^{1/2}/\mathcal{L}_{i-1},  
\end{equation}
where the determinants $\mathcal{L}$ and $\mathcal{M}$ are defined by 
\be
\label{eqn:det1}
\mathcal{L}_{n}=\left|\begin{array}{ccccc} 1&\mu_1&\mu_2&\cdots&\mu_{n}\\
\mu_1&\mu_2&\mu_3&\cdots&\mu_{n+1}\\ &\vdots&&&\vdots\\
\mu_{n}&\cdots&&&\mu_{2n}\\
\end{array}
\right|
\ee
and 
\be
\label{eqn:det2}
\mathcal{M}_{n}=\left|\begin{array}{cccccc}
1&\mu_1&\mu_2&\cdots&\mu_{n-1}&\mu_{n+1}\\
\mu_1&\mu_2&\mu_3&\cdots&\mu_{n}&\mu_{n+2}\\
&\vdots&&&\vdots&\vdots\\
\mu_{n}&\cdots&&&\mu_{2n-1}&\mu_{2n+1}\\
\end{array}
\right|,
\ee
with $\mu_\lambda \equiv \la v_1|H^\lambda| v_1\ra$,
and with $\mathcal{M}_{-1}$=0, $\mathcal{M}_0=\mu_1$,
and $\mathcal{L}_0$=1.  The
NMM uses these equations to determine the
Lanczos matrix $L(n,|v_1(y)\ra$ from the $\mu_\lambda(y)$.   Then one can proceed in the usual
way to diagonalize this matrix and then construct the
strength function, using Eq.~\ref{eqn:strengthdistribution}. 

While sound mathematically, the NMM is problematic numerically
because the solution to the classical moments problem embodied
in Eqs.~\ref{eqn:alpha} and ~\ref{eqn:beta} is not a stable one.
For example, Whitehead and Watt provide in Ref.~\cite{whit3}
a simple 4 $\times$ 4 matrix example in which significant loss
of accuracy occurs.There have been attempts \cite{fletcher91,nair82} to finite alternate
methods that use moments in a manner that will improve the
inversion to a distribution.  None of these have proven to
have the stability of a direct Lanczos construction, however. 
In the numerical tests we performed,
significant errors typically occurred for $n \sim 20$, in calculations
performed with 64-bit accuracy.  As is apparent from many of
the calculations presented in this paper, often 50-200 iterations
are required before a response function is accurately reconstructed,
for resolutions we explored.

\subsection{Legendre Moments Method}

Operationally, the rapid loss of precision in the NMM
occurs because of severe cancellations in the determinants
(Eqs.~\ref{eqn:det1} and \ref{eqn:det2}) that arise from the
dominance of the largest eigenvalues in the highest-order moments.
This dominance occurs because naive moments methods work with a set of
non-orthogonal basis vectors: $H^n|v_1\ra$ with $n$ non-negative
integers.  The Lanczos algorithm, on the other hand, builds a set of
orthogonal vectors $|v_n\ra$ as it goes, each carrying information
about $H$ that is independent from that in its predecessors, and this
produces the great stability of the algorithm.

We attempted to find a new way to iterate for $\alpha_i$ and $\beta_i$
in terms of moments that, like the original Lanczos algorithm, are
built on a set of orthogonal vectors, $|w_i\ra$, constructed as part
of the iteration.  We describe a first attempt at an iterative method,
which fails because it is equivalent to the NMM.  We then give a
closely analogous procedure based on linear combinations of moments
that proved, at times, to be considerably more stable.

We begin as in the NMM procedure by calculating the mixed moments
$\Omega^\lambda_{ij}$, from which we can then evaluate the moments
$\langle v_1(y)| H^m |v_1(y)\rangle$ as a function of $y$.

We then build up the orthogonal vectors $|w_i\ra$ of dimension
$i$ iteratively, starting with 
\be
\label{alphas}
 |w_1\ra=\left(\al_1\right) 
 \ee 
\be
|w_2\ra=\left(\begin{array}{c} \al_1 \\ \bt_1 \end{array} \right), \ee
where $\alpha_1$, $\beta_1$ are easily identified from the moments
$\la v_1|H|v_1\ra=\al_1$ and $\la v_1|H^2|v_2\ra-\al_1^2=\bt_1^2$
(defined as in ordinary Lanczos).  With these vectors in hand we
compute the next $\alpha_i$, $\beta_i$, and $|w_n\ra$ from
\begin{eqnarray}
\alpha_n=\frac{\la v_1|H^{2n-1}|v_1\ra-\la
  w_{n-1}|L_{n-1}|w_{n-1}\ra}{\bt_1^2\bt_2^2\cdots\bt_{n-1}^2}\nonumber \\
-2(\al_1+\al_2\cdots+\al_n)
\label{alphmom}
\end{eqnarray}
\be
\bt_{n-1}^2=\frac{\la v_1|H^{2n-2}|v_1\ra-\la
  w_{n-1}|w_{n-1}\ra}{\bt_1^2\bt_2^2\cdots\bt_{n-2}^2}
\label{betmom}
\ee
\be
|w_n\ra=L_n\left(\begin{array}{c}|w_{n-1}\ra \\
    \bt_1\bt_2\cdots\bt_{n-1} \end{array}\right),
\label{eqn:wmom}
\ee
where $L_n$ is the truncated Lanczos matrix of Eq.~\ref{eqn:lanczosmatrix}.

In this way we construct the Lanczos matrix, which can then be
diagonalized and used to derive the response function in the usual
way.  This algorithm yields significant stability improvements over
the procedure outlined in the NMM discussion.
However, the new algorithm still suffers
from cancellations between terms in both Eqs.~\ref{alphmom} and
\ref{betmom}.  Successively higher moments yield increasingly large
numbers on each side of the minus signs in those equations.  These
numbers are subtracted to yield successive $\al_i$ and $\bt_i$ that
remain of order unity.  Such large cancellations lead to loss of
precision that, though less severe than in naive moments methods, has
the same root and the same consequence of failure in that eventually
$\beta_n^2<0$.  At this point the algorithm fails.

One possible solution to this problem is to find combinations of moments that
remain relatively stable in size, e.g., some set of orthogonal
polynomials.  If we scale and shift energies so that the range of
eigenvalues can be mapped onto [-1,1], one obvious choice would be
Legendre polynomials.  They contain the same information as the
moments, and their matrix elements will have the same dependence in
$y$.  However, as they have a fixed magnitude at the endpoints, they
are not overwhelmed by the extremal eigenvalues for large $n$.  In the
LMM the moments of the NMM are replaced by
\begin{eqnarray*}
\la v_1|\hat{P}_0^2|v_1 \ra& \equiv& \la q_0|q_0\ra \\
\la v_1|\hat{P}_1^2|v_1 \ra&\equiv&\la q_1|q_1\ra \\
\la v_1|\hat{P}_1H\hat{P}_1|v_1 \ra&=&\la q_1|H|q_1\ra \\
\la v_1|\hat{P}_2^2|v_1 \ra&\equiv&\la q_2|q_2\ra \\
\la v_1|\hat{P}_2H\hat{P}_2|v_1 \ra&=&\la q_2|H|q_2\ra \\
&\vdots&
\end{eqnarray*}
The operators $\hat{P}_l(H)$ are linear combinations of powers of $H$,
with coefficients identical to those on the corresponding powers of
the scalar $x$ in the Legendre polynomials $P_l(x)$.  Instead of
computing moments, we calculate the overlaps $\la q_l|q_l\ra$, using
the Legendre polynomial recurrence relation to generate $|q_l\ra$.
The vectors $|q_l\ra$ have a simple expansion in $y$ because they are
built from linear combinations of the $|v_1^i\rangle$, so \be
|q_l\ra=|q_l^1\ra+|q_l^2\ra y+\cdots|q_l^n\ra y^{n-1}, \ee and
$|q_l^i\rangle$ is easily computed from $|v_1^i\rangle$ at the start
of the calculation.  The recurrence relations used are then
\begin{eqnarray}
  |q_0^i\rangle & = & |v_1^i\rangle\\
  |q_1^i\rangle & = & H |v_1^i\rangle  \\
|q_{l+1}^i\rangle & =& 
\frac{2l+1}{l+1}H|q_l^i\rangle - \frac{l}{l+1} |q_{l-1}^i\rangle.
\end{eqnarray}

It is straighforward to find the Lanczos matrix $L_n$ in terms of the
$\langle q_n|q_n\rangle$.  This resulting inversion -- from Legendre
moments to the tridiagonal matrix -- proved to be significantly more
stable than the NMM inversion.  The recurrence relations to compute
$L_n$ are

\begin{eqnarray}
\alpha_n=\frac{\la q_{n-1}|H|q_{n-1}\ra-\la
  w_{n-1}|L_{n-1}|w_{n-1}\ra}{\bt_1^2\bt_2^2\cdots\bt_{n-1}^2}\nonumber \\
-2\left(\frac{(2n-3)!!}{(n-1)!}\right)^2(\al_1+\al_2\cdots+\al_n)
\label{eqn:alphalm}
\end{eqnarray}
\be
\label{eqn:betalm}
\bt_{n-1}^2=\left(\frac{(n-1)!}{(2n-3)!!}\right)^2\frac{\la q_{n-1}|q_{n-1}\ra-\la
  w_{n-1}|w_{n-1}\ra}{\bt_1^2\bt_2^2\cdots\bt_{n-2}^2},
\ee
\begin{eqnarray}
|w_n\ra=\left(\frac{2n-3}{n-1}\right)L_n\left(\begin{array}{c}|w_{n-1}\ra \\ 0
  \end{array}\right)\nonumber \\
-\left(\frac{n-2}{n-1}\right)\left(\begin{array}{c}|w_{n-2}\ra
    \\ 0 \\ 0 \end{array}\right) 
\label{eqn:wlm}
\end{eqnarray}
with $|w_1\rangle$, $|w_2\rangle$, $\alpha_1$, and $\beta_1$ as
before.

\begin{figure}
\includegraphics[width=9.0cm]{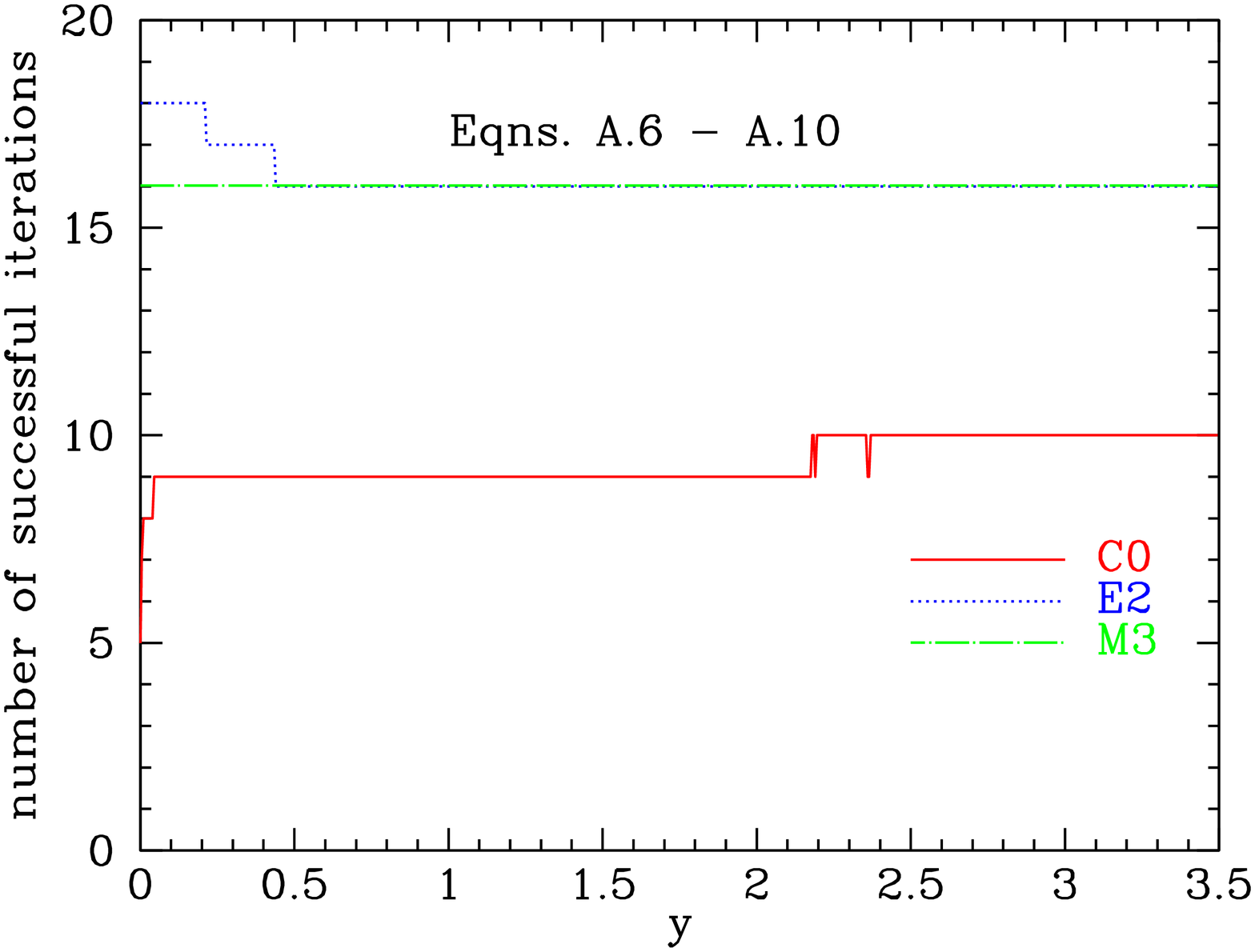}
\includegraphics[width=9.0cm]{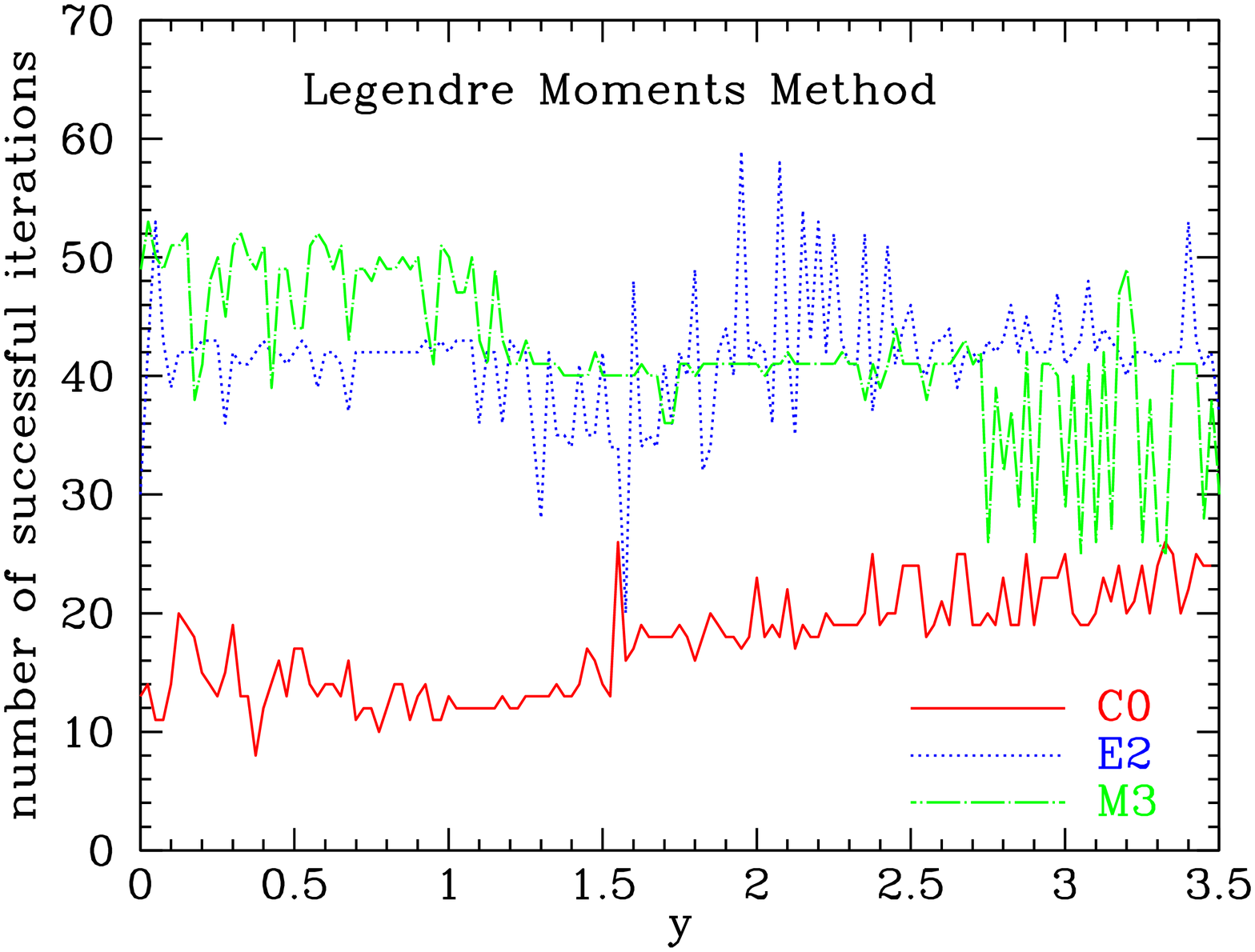}
\caption{The number of iterations
before loss of precision produces $\beta_i<0$ in the
moments algorithms, for each operator computed in our $^{28}$Si
example case, as functions of momentum transfer.  The upper
graph is the result for the method of Eqs.~\ref{alphas}--\ref{eqn:wmom} , while the
bottom graph gives the Legendre Moments
Method (Eqs.~\ref{eqn:alphalm}--\ref{eqn:wlm}).}
\label{fig:stability}
\end{figure}

These two moments methods (Eqs.~\ref{alphmom}--\ref{eqn:wmom} and
Eqs.~\ref{eqn:alphalm}--\ref{eqn:wlm}) are defined by similar
procedures.  We compare their stability in Fig. 10.
In our experience, the Legendre moments algorithm can sometimes run to
very high numbers ($> 80$) of iterations with no significant loss of
precision, but it also sometimes runs aground quickly on cancellations
in Eq.~\ref{eqn:betalm}.  The lack of predictability is clearly an issue.

Our failure to identify a more reliable method for determining distributions
from moments motivated us to look for other approaches, e.g., ones that might not
capture exact information on the moments at every $y$ or
exactly guarantee positive-definiteness of the response function, but would
remain stable and accurate under continued iteration.  This
led to the PMM method we favor, as well as one other moment-preserving
approach described below.

\subsection{Legendre Coefficients Method} 

Rather than  constructing a Lanczos
matrix, diagonalizing, and  finding the strength distribution, an
approximate strength distribution can be computed directly
from an expansion in Legendre polynomials, 
\be
S(\omega,y)=y^{J-K}e^{-2y}\sum_{l=0}^\infty a_l(y)P_l(\omega).  
\ee 
The coefficients $a_l(y)$
can be computed using the orthogonality relation for Legendre
polynomials and the identity 
\be 
y^{J-K}e^{-2y} |c(y)|^2\la
v_1|\hat{P}_l(H)|v_1\ra=\int S(\omega,y) P_l(\omega)d\omega, 
\ee 
so that 
\be
a_l(y)=\frac{2l+1}{2}|c(y)|^2\la v_1|\hat{P}_l(H)|v_1\ra.  
\ee 
We
refer to the method of computing response functions from Legendre
coefficients computed with these relations as the ``Legendre
coefficients'' method.

Here again, $H$ must be shifted and scaled to the interval
$\left[-1,1\right]$ before calculating these matrix elements, then
shifted and scaled back at the end to obtain the true strength
distribution.  A numerically stable approach to calculating the
$a_l(y)$ uses the same vectors $|q_n\ra$ defined for the
Legendre moments method, constructed in the same way, so that 
\be
a_l(y)=\frac{2l+1}{2}\sum_{j,k=0}^m c_jc_ky^{j+k}\la v_1^k|q_l^j\ra, 
\ee
where $c_j$ are defined in Eq.~\ref{eqn:again}.

The approximate strength function $S_{LC}^n$ represented by the first
$n$ Legendre polynomials will always contain oscillations about the
true strength function, at an energy scale given by the order of the
highest Legendre polynomial in the expansion, so roughly on the scale
$\Delta \omega/n$, where $\Delta \omega$ is the difference between the
maximum and minimum eigenvalues of $H$.  (Since the true response
function is a sum of delta functions, its expansion in Legendre
polynomials never truncates.)  There is no restriction on the sign of
$S_{LC}^n(\omega,y)$, so its oscillations may make it negative in
places even though $S(\omega,y)$ is strictly positive.  For these
reasons, practical use of this method would probably require smoothing
smoothing the function $S_{LC}^n(\omega,y)$ to finite resolution over
scales of approximately $\Delta \omega/n$.  Since we have a Legendre
expansion of $S_{LC}^n$, the convolution to produce the smoothed
function may be performed efficiently (i.e., reduced to a series of
matrix multiplications) by working with Legendre expansions of the
smoothing function.  The Legendre coefficient expansion may be useful
in applications where the response function needs to be convolved with
some other function, for example, a thermal energy distribution.  On
the other hand, while the Lanczos methods converge more rapidly at the
ends of the eigenvalue spectrum than in the middle, that is not in
general the case for the Legendre expansion.

We also note that, whereas the ordinary Lanczos algorithm and our
variants on it need to run $n$ iterations on each vector piece to
reproduce $2n-1$ moments of the response function, the Legendre
coefficients methods needs to run $2n-1$ iterations to produce $2n-1$
moments, essentially twice as long as the other methods in this
Appendix.  As in the Lanczos method, each iteration contains as
its basic time-consuming step a matrix-vector multiplication in the
original large basis.

It has come to our attention that a closely-related method is in use
in physical chemistry, where it is used to compute the
quantum-mechanical time evolution of molecular states
\cite{wall95,kosloff87}.  Authors working in that area point out that
if Chebyshev polynomials are used instead of Legendre polynomials, the
vectors $|q_l\rangle$ may still be found by recurrence, but only the
first $n$ such vectors need to be computed in order to find the first
$2n$ terms of the expansion in Chebyshev polynomials.


\begin{thebibliography}{000}

\bibitem{caur}  E. Caurier, G. Martinez-Pinedo, F. Nowacki, A. Poves,
and A. P. Zuker, nucl-th/0402046/ (Rev. Mod. Phys., in press)

\bibitem{brown} B. A. Brown and B. H. Wildenthal, Ann. Rev. Nucl.
Part. Sci. {\bf 38}, 29 (1988); S. Cohen and D. Kurath, Nucl. Phys.
{\bf 73}, 1 (1965.

\bibitem{song} W. C. Haxton and C. L. Song, Phys. Rev. Lett. {\bf 84},
5484 (2000); P. Navratil, J. P. Vary, and B. R. Barrett, Phys. Rev. Lett.
{\bf 84}, 5728 (2000); S. Bogner, T. T. S. Kuo, L. Coraggio, A. Covello,
and N. Itaco, Phys. Rev. C{\bf 65}, 051301 (2002); T. S. Luu, S. Bogner,
W. C. Haxton, and P. Navratil, Phys. Rev. C.{\bf 70}, 014316 (2004).

\bibitem{lanczos} C. Lanczos, J. Res. Nat. Bur. Stand. {\bf 45}, 255 (1950);
J. H. Wilkinson, {\it The Algebraic Eigenvalue Problem} (Clarendon Press,
Oxford, 1965).

\bibitem{langanke00} K. Langanke and G. Martinez-Pinedo, Nucl.
Phys. A{\bf 673}, 481 (2000); E. Caurier, K, Langanke, G. Martinez-Pinedo,
and F. Nowacki, Nucl. Phys. A{\bf 653}, 439, 1999.

\bibitem{heger01} A. Heger, S. E. Woosley, G. Martinez-Pinedo, and K.
Langanke, Astrophys. J. {\bf 560}, 307 (2001).

\bibitem{hix03} W. R. Hix, O. E. B. Messer, A. Mezzacappa, M. Liebendoerfer,
J. Sampaio, K. Langanke, D. J. Dean, and G. Martinez-Pinedo, Phys. Rev.
Lett. {\bf 91}, 201102 (2003).

\bibitem{haxton88} W. C. Haxton, Phys. Rev. Lett. {\bf 60}, 1999 (1988);
S. E. Woosley, D. H. Hartmann, R. D. Hoffman, and W. C. Haxton,
Astrophys. J. {\bf 356}, 272 (1990).

\bibitem{cullum85} J. Cullum and R. Willoughby, {\it Lanczos Algorithms
for Large Symmetric Eigenvalue Computations}, vol, 1 (Birkh\"{a}user, Boston,
1985)

\bibitem{whit1} R. R. Whitehead, in {\it Moment Methods in Many Fermion Systems},
ed. B. J. Dalton, S. M. Grimes, J. P. Vary, and S. A. Williams (Plenum Press,
New York, 1980), p. 235.

\bibitem{whit2} R. R. Whitehead {\it et al.}, Adv. in Nucl. Phys. {\bf 9}, 123 (1977).

\bibitem{haydock} R. Haydock, J. Phys. {\bf A7}, 2120 (1974); R. Haydock, in
{\it Computational Methods in Classical and Quantum Physics}, ed. M. B.
Hooper (Advance Publications, London, 1976), p. 268.

\bibitem{engel} J. Engel, W. C. Haxton, and P. Vogel, Phys. Rev. C{\bf 46},
2153 (1992).

\bibitem{walecka} T. DeForest, Jr., and J. D. Walecka, Adv. Phys. {\bf 15}, 1
(1966); J. D. Walecka, in {\it Muon Physics}, ed. V. W. Hughes and
C. S. Wu (Academic Press, New York, 1975), p. 113.

\bibitem{donnelly} T. W. Donnelly and W. C. Haxton, Atomic Data and
Nucl. Data Tables {\bf 23}, 103 (1979).

\bibitem{Marchisio} There have been applications of Lorentz transform methods
to few-body nuclei, including one that employed Lanczos techniques, though the
approach and motivations differ from those under discussion here.  See, for
example, M. A. Marchisio, N. Barnea, W. Leidemann, and G. Orlandini,
Few-Body Syst. {\bf 33}, 259 (2003) and nucl-th/0202009/.

\bibitem{akhiezer} N. Akhiezer, {\it The Classical Moment Problem} (Oliver
and Boyd, Edinburgh,1965).

\bibitem{whit3} R. R. Whitehead and A. Watt, J. Phys. G {\bf 4}, 835
(1978) and J. Phys. A {\bf 14}, 1887 (1981).

\bibitem{fletcher91} J. Fletcher, J. Phys. A {\bf 24}, L7 (1991).

\bibitem{nair82} C. Radhakrishnan and S. Nair, Phys. Lett. {\bf 91A}, 64 (1982).

\bibitem{wall95} M. R. Wall and D. Neuhauser, J. Chem. Phys. {\bf 102}, 8011 (1995).

\bibitem{kosloff87} R. Kosloff, J. Phys. Chem. {\bf 92}, 2087 (1987).

\end{thebibliography}
\end{document}